\newif\ifpaper
\papertrue 
\ifpaper
\documentclass[10pt,journal,final,twoside,twocolumn]{IEEEtran}
\else
\documentclass[journal,12pt,onecolumn,draftclsnofoot,finalsubmission]{IEEEtran}
\fi

\usepackage[utf8]{inputenc}

\usepackage{amsmath, bbm}
\usepackage{amssymb}
\usepackage{amsfonts}
\usepackage{amsthm}
\usepackage{algorithm}
\usepackage{algorithmic}
\usepackage{nicefrac}
\usepackage{bm}
\usepackage{hyperref}
\usepackage{cite}
\usepackage[utf8]{inputenc} 
\usepackage[T1]{fontenc}
\usepackage{graphicx}
\usepackage[font=small]{caption}
\usepackage{tikz}
\usetikzlibrary{positioning}
\usetikzlibrary{arrows}
\usetikzlibrary{matrix}
\usetikzlibrary{backgrounds}
\usepackage{pgfplots}
\usepackage{epstopdf}
\usepackage[export]{adjustbox}
\usepackage[labelformat=simple]{subcaption}
\usepackage{multirow}
\usepackage{lipsum}
\usepackage{mathtools}
\usepackage{longtable}

\IEEEoverridecommandlockouts
\usepackage{etoolbox}
\makeatletter
\patchcmd{\@makecaption}
{\scshape}
{}
{}
{}
\makeatletter
\patchcmd{\@makecaption}
{\\}
{.\ }
{}
{}

\newtheorem{Problem}{Problem}

\input{mysymbol.sty}

\long\def\comment#1{}

\newcommand \COMMENTS[1] {{\color{gray}\textit{/* #1 */}}}

\title{Distributed Link Sparsification for Scalable Scheduling Using Graph Neural Networks}

\author{Zhongyuan Zhao,~\IEEEmembership{Member,~IEEE,}
        Gunjan Verma,~\IEEEmembership{Member,~IEEE,}\\
        Ananthram Swami,~\IEEEmembership{Life~Fellow,~IEEE,}
        and Santiago Segarra,~\IEEEmembership{Senior~Member,~IEEE}
	\thanks{
    Received 27 February 2024; revised 20 January 2025, 17 June 2025, and 13 August 2025; accepted 1 September 2025. Date of publication DD MMM 2026; date of current version DD MMM 2026.
Research was sponsored by the DEVCOM ARL Army Research Office and was accomplished under Cooperative Agreement Number W911NF-19-2-0269. 
		The views and conclusions contained in this document are those of the authors and should not be interpreted as representing the official policies, either expressed or implied, of the Army Research Office or the U.S. Government. 
		The U.S. Government is authorized to reproduce and distribute reprints for Government purposes notwithstanding any copyright notation herein.}
	\thanks{Zhongyuan Zhao and Santiago Segarra are with the Department of Electrical and Computer Engineering, Rice University, USA. e-mails: \{zhongyuan.zhao, segarra\}@rice.edu}
	\thanks{Gunjan Verma and Ananthram Swami are with the DEVCOM Army Research Laboratory (ARL), USA. e-mails: \{gunjan.verma.civ, ananthram.swami.civ\}@army.mil.}
	\thanks{Preliminary results were presented in~\cite{zhao2022icassp_b}.}
	 \thanks{Digital Object Identifier 10.1109/TWC.2025.3606741}
}

\IEEEaftertitletext{\vspace{-1\baselineskip}}

\pgfplotsset{compat=1.16}
\begin{document}
\markboth{IEEE Transactions on Wireless Communications, Vol. XX, No. YY, September 2025}%
{Zhao \MakeLowercase{\textit{et al.}}: Distributed Link Sparsification for Scalable Scheduling Using Graph Neural Networks}

\maketitle

\begin{abstract}
In wireless networks characterized by dense connectivity, the significant signaling overhead generated by distributed link scheduling algorithms can exacerbate issues like congestion, energy consumption, and radio footprint expansion.
To mitigate these challenges, we propose a distributed link sparsification scheme employing graph neural networks (GNNs) to reduce scheduling overhead for delay-tolerant traffic while maintaining network capacity. 
A GNN module is trained to adjust contention thresholds for individual links based on traffic statistics and network topology, enabling links to withdraw from scheduling contention when they are unlikely to succeed.
Our approach is facilitated by a novel offline constrained {unsupervised} learning algorithm capable of balancing two competing objectives: minimizing scheduling overhead while ensuring that total utility meets the required level.
In simulated wireless multi-hop networks with up to 500 links, our link sparsification technique effectively alleviates network congestion and reduces radio footprints across four distinct distributed link scheduling protocols.
\end{abstract}
\section{Introduction}\label{sec:intro}
The proliferation of wireless devices and emerging machine-type communications (MTC) \cite{cisco2020} has led to new requirements for next-generation wireless networks, including massive access in ultra-dense networks, spectrum and energy efficiencies, multi-hop connectivity, and scalability \cite{kott2016internet,akyildiz20206g,chen2021massive,Sharma2020toward}.
A promising solution to these challenges is self-organizing wireless multi-hop networks, which have been applied to scenarios where infrastructure is infeasible or overloaded, such as military communications, satellite communications, vehicular/drone networks, Internet of Things (IoT), and 5G/6G (device-to-device (D2D), wireless backhaul, integrated access and backhaul (IAB))~\cite{Lin06,sarkar2013ad,kott2016internet,Patriciello2016,akyildiz20206g,chen2021massive,Sharma2020toward,Cudak2021}.
In ultra-dense cellular IoT networks~\cite{Sharma2020toward}, delay-tolerant MTC  could be enabled by using cluster heads that first aggregate data from their nearby IoT devices based on distributed clustering and link scheduling~\cite{zhao2023iclr,zhao2022twc}, then relay data to the base-station either directly or over multiple hops.
Mobile MTC, such as vehicular networks and Internet of Battlefield Things~\cite{kott2016internet}, can also benefit from the congestion management of self-organizing wireless multi-hop networks. 
The key enabler of self-organizing wireless multi-hop networks is distributed resource allocation.
This includes link scheduling, which determines which links should transmit and when they should transmit \cite{Joo09,marques2011optimal}.
In this work, we focus on link scheduling in wireless multi-hop networks with orthogonal multiple access (OMA).
Compared to non-orthogonal multiple access (NOMA) \cite{chen2021massive}, OMA requires less sophisticated transceivers and coordination, therefore, is more friendly to MTC of low-end devices in shared spectrum~\cite{Sharma2020toward} or harsh/hostile environments~\cite{kott2016internet}.
In 5G/6G, OMA is also better suited for D2D, wireless backhaul, and IAB~\cite{akyildiz20206g,Cudak2021}. 

\IEEEpubidadjcol

In wireless multi-hop networks with OMA, link scheduling is typically formulated as solving a maximum weighted independent set (MWIS) problem on a conflict graph~\cite{basagni2001finding,Joo09,cheng2009complexity,joo2012local,joo2015distributed,marques2011optimal,du2016new,paschalidis2015message,joo2010complexity,zhao2021icassp,zhao2022twc}.
In a conflict graph, a vertex represents a link in the wireless network, an edge captures the interference relationship between two links, and the vertex weight is the utility of scheduling the corresponding link.
Known to be NP-hard~\cite{cheng2009complexity,joo2010complexity}, the MWIS problem is solved approximately in practical schedulers. 
Common distributed heuristics include distributed greedy MaxWeight schedulers \cite{joo2012local,joo2015distributed}, message passing algorithms \cite{paschalidis2015message,du2016new}, their hybrids \cite{zhao2021icassp,zhao2022twc}, and randomized algorithms like carrier sense multiple access (CSMA) \cite{ni2012qcsma,jiang2010distCSMA,Lin2009constant}, 
many of which are inspired by Luby's algorithm~\cite{luby1985simple}.

The overhead of distributed link scheduling, referred to as the signaling overhead in MaxWeight schedulers~\cite{joo2012local,joo2015distributed,zhao2021icassp,zhao2022twc} and collision rates and idle listening in CSMA~\cite{ni2012qcsma,jiang2010distCSMA,Lin2009constant}, increases with the number of contending neighbors in the network.
In dense networks characterized by large contention neighborhoods, the scheduling overhead can be substantial, reducing network capacity~\cite{cisco2020,chen2021massive} while increasing congestion, instability, energy consumption, radio footprint, and security vulnerability in network operations~\cite{chen2021massive,Testi2021blind,ye2004coordinated,Santi2005topology}.
While existing approaches like topology control by adjusting antenna directivity and transmit power \cite{Ramanathan2004ch5,Santi2005topology,ray2016hybrid}, sleep scheduling \cite{ray2016hybrid,ye2004coordinated,guha2011greenwave,long2020collaborative}, cross-layer optimization \cite{lin2010lowcomplexity,xiang2014energy,wu2020energy}, and network coding~\cite{Ahlswede2000} can partially mitigate challenges of dense wireless networks, they often fall short when dealing with the specific issue of massive access. 

In this work, we seek scalable scheduling schemes for delay-tolerant traffic to reduce the scheduling overhead in dense wireless networks.
Specifically, we introduce an additional step of \emph{link sparsification} into distributed link scheduling to limit the number of participants in the transmission contention by filtering out links unlikely to win, such as those with utilities below certain thresholds~\cite{zhao2022icassp_b}.
A \emph{baseline} approach is to set a global threshold {$u^{(\eta)}$} across the network based on the empirical cumulative distribution function (eCDF) of the per-link utility and a prescribed cut-off quantile {$\eta$}.
On an arbitrary link $v$, such a baseline sparsification can be expressed as 
\begin{equation}\label{E:baseline}
    h\big(u(v,t)\big)=u(v,t)H\left(u(v,t)-u^{(\eta)}\right)\;,
\end{equation}
where $u(v,t)\geq 0$ is the utility of scheduling link $v$ at time $t$ and $H(\cdot)$ is the Heaviside (step) function.
If $u(v,t)\leq u^{(\eta)}$, $h\big(u(v,t)\big)=0$, and link $v$ is muted for time step $t$ without contention so as to reduce scheduling overhead.
Since link utility is usually a function of queue length \cite{joo2012local,joo2015distributed} or sojourn time \cite{hai2018delay},
every link will eventually contest for scheduling as the backlog or waiting time grows.
The increasing bandwidth of air-interfaces~\cite{cisco2020} and the capacity of on-chip memory make it feasible to trade off some latency for more connectivity by transmitting larger data payloads less frequently.

\vspace{2mm}
\noindent
\emph{Can we improve upon this (statistical) baseline?}
Ideally, we could refine the global threshold in~\eqref{E:baseline} for each link based on its local traffic and topological information. 
For instance, lowering the thresholds for links with fewer contending neighbors, and vice versa, can increase transmission opportunities in areas with lower scheduling overhead while reducing contentions in denser areas, potentially leading to improved resource efficiency in the network.
Technically, such refinement can be implemented using a vector $\bbz\in\reals_{+}^{|\ccalV|}$, which scales the global threshold individually for each link in the network $v\in\ccalV$,
\begin{equation}\label{E:param}
h_v\big(u(v,t);\bbz\big) = u(v,t) \, H\left(u(v,t)- z(v) u^{(\eta)}\right)\;,
\end{equation}
where $z(v)=\bbz_v$ is the $v$th element of vector $\bbz$.
Vector $\bbz$ can be optimized to produce a policy that is Pareto superior to the statistical baseline, achieving at least the same total utility while reducing scheduling overhead across the network.

To achieve this goal, we propose to train a GNN to generate topology-aware multipliers $\bbz$ for the localized link sparsification $\{h_v(\cdot)\}_{v\in\ccalV}$ in~\eqref{E:param}.
Compared to conventional optimization methods or deep neural networks (DNNs), GNNs have several unique advantages:
1)~GNNs are designed to process relational data, including the conflict relationships between wireless links, which serve as key inputs for threshold refinement.
2)~GNNs can operate in a fully distributed manner via message passing between neighboring nodes, eliminating the need to aggregate the full network state at a centralized server, which is critical to scalability and robustness.
3)~The built-in permutation equivariance of GNNs aligns naturally with tasks in wireless networks--the output of GNNs remains consistent under permutations of input node ordering-- making them more efficient in data and structure compared to DNNs, which must learn such symmetry from data.
4)~GNNs also generalize well to various network topologies~\cite{wang2024generalization}, allowing a trained GNN to seamlessly scale to larger networks and adapt to node mobility in wireless networks.

However, applying GNNs to link sparsification faces unique challenges.
First, GNNs' reliance on message passing between neighboring devices inherently conflicts with our very purpose of minimizing signaling overhead, making it counterproductive to frequently call GNNs~\cite{wang2022learning}. 
Our solution enables GNNs to operate infrequently and asynchronously.
Another challenge is to design a training scheme that can train GNNs in non-differentiable pipelines, similar to~\cite{zhao2021icassp,zhao2022twc,zhao2023iclr}, while optimizing two competing objectives, the total utility and scheduling overhead.
To overcome these challenges, we propose a novel, first-order constrained unsupervised learning algorithm that can outperform the standard approach of zeroth-order optimization (ZOO)~\cite{liu2020primer}.
Although the training is offline and centralized in simulated environments, the trained GNNs and the sparse schedulers can be deployed to infrastructureless networks and executed in a fully distributed manner.

\vspace{1mm}
\noindent
{\bf Contribution.}  
The major contributions of this paper are:
\begin{itemize}
    \item We propose the first GNN-based distributed link sparsification for scalable link scheduling in dense networks by exploiting the topology of the interference graph.
    \item We develop alternating stochastic gradient descent (Alt-SGD), a first-order constrained {unsupervised} learning algorithm that can train the GNNs in a non-differentiable pipeline while balancing two competing objectives.
    \item Through simulation, we demonstrate that our GNN-based threshold policy can outperform the statistical baseline in dense wireless networks with four different scheduling protocols, and can generalize across topologies.
    \item We provide practical guidelines on implementing  contention and conflict graph construction simultaneously, asynchronous operations for GNNs, gathering the eCDF of per-link utility, and setting the global cut-off quantile. 
\end{itemize}

\noindent
{\bf Notation:} 
We adopt the following notational convention: 
$ (\cdot)^\top $, $ \odot $, and $ |\cdot| $ respectively represent transpose operator, Hadamard (element-wise) product operator, and the cardinality of a set.
$ \mathbbm{1}(\cdot) $ is the indicator function, and $ \mathbb{E}(\cdot) $ stands for expectation.
Upright bold lower-case symbol, e.g., $\bbx$, denotes a column vector, and $\bbx_i$ denotes the $i$th element of vector $\bbx$. 
Upright bold upper-case symbol $\bbX$ denotes a matrix, of which the element at row $i$ and column $j$ is denoted by $\bbX_{ij}$, the entire row $i$ by $\bbX_{i*}$, and the entire column $j$ by $\bbX_{*j}$.
The major notations, though not exhaustive, are described in Table~\ref{tab:symbols}.

\section{Related Work}
\subsection{Scalable Resource Allocation}
Scalable link scheduling has been a long-standing challenge for wireless multi-hop networks, where centralized scheduling is hindered by the high communication overhead of gathering the full network state to a server.
Distributed scheduling ~\cite{basagni2001finding,Joo09,cheng2009complexity,Lin2009constant,ni2012qcsma,joo2012local,joo2015distributed,du2016new,paschalidis2015message,joo2010complexity,zhao2021icassp,zhao2022twc} addresses the scalability concerns by tackling the MWIS problem (or its dual, maximum weighted matching) through parallel local contentions among interfering neighbors.
These contentions can take either deterministic forms, relying on exchange and comparison of utilities among neighboring links \cite{joo2012local,joo2015distributed,zhao2021icassp,zhao2022twc}, or randomized forms, employing weighted CSMA \cite{ni2012qcsma,Lin2009constant,huang2012low,gupta2009low}.
Classical distributed schedulers~\cite{joo2012local,joo2015distributed} are capable of solving the MWIS problem in $\ccalO(\log |\ccalV|)$ rounds of local message exchange, where $|\ccalV|$ is the number of links in the network.
Scheduling within $\ccalO(1)$ (constant) rounds of local exchange can be achieved by truncating the iterations in \cite{joo2012local,joo2015distributed} or by employing one round of local exchange followed by weighted CSMA \cite{ni2012qcsma,Lin2009constant,huang2012low,gupta2009low}.
However, for ultra-dense networks, the reduction in rounds of local exchanges is insufficient, since the overhead of each round of local exchange (or the collision rate in CSMA) increases with the conflict degree (the number of interfering neighbors), potentially reaching prohibitively high levels.
In contrast to these approaches, our link sparsification scheme lowers the scheduling overhead by reducing the conflict degree.
While certain strategies in our approach, such as \emph{buffer-and-transmit}~\cite{Ryu2012timescale} and probabilistic participation in contention~\cite{Lin2009constant} are not novel,
our method distinguishes itself by eliminating the need for message exchange across the entire neighborhood at every time step, a requirement present in prior models~\cite{Lin2009constant,Ryu2012timescale}.

Besides scheduling, other resource allocation schemes can also help reduce the signaling overhead in wireless multi-hop networks.  
Topology control~\cite{Ramanathan2004ch5,Santi2005topology,ray2016hybrid} seeks to maintain a connected topology of the network with minimum energy consumption by adjusting antenna beams and/or transmit power.
In dense networks, topology control can reduce the scheduling overhead by limiting the neighborhood size, at the cost of increased overhearing, latency, and risks of congestion at some critical nodes. 
Sleep scheduling \cite{ray2016hybrid,ye2004coordinated,guha2011greenwave,long2020collaborative} puts devices with very low traffic demand into periodic sleep mode in a standalone or coordinated manner.
It mainly aims to extend the battery lifetime of wireless devices, but can also reduce the densities of connectivity and interference at a given moment.
In comparison, link sparsification can benefit active devices with higher traffic demands.
Cross-layer optimization \cite{lin2010lowcomplexity,xiang2014energy,wu2020energy} jointly optimizes power control, link scheduling, and routing to improve energy efficiency and latency, but it generally entails similar or higher signaling overhead than the aforementioned MWIS schedulers, and may also require centralized computing~\cite{xiang2014energy,wu2020energy}.
Network coding~\cite{Ahlswede2000} seeks efficient utilization of network capacity by aggregating data packets from different sources destined to the same sink, rather than forwarding them separately, which can also reduce the level of contention.
However, the aforementioned methods by themselves are generally insufficient for scalable resource allocation in ultra-dense networks, whereas link sparsification can \emph{complement} them. 
Lastly, queueing systems with threshold policy have been studied for decades~\cite{yadin1963queueing,kella1989threshold} and applied to energy saving for wireless sensor/IoT networks~\cite{lee2013n,Qi2020traffic}.
However, these steady state analyses of queueing systems are not applicable to link scheduling with queues subject to interference constraints, where the service rates (link rates) are influenced by the thresholds through scheduling overhead rather than following a stationary distribution.

\subsection{Machine Learning in Wireless Networks}
Machine learning (ML) for wireless networks faces four major challenges.
Firstly, while data is abundant in real or simulated systems, labeling it is often difficult~\cite{Sharma2020toward}. 
Consequently, learning without labeling is often preferred over supervised and semi-supervised learning in wireless systems. 
Secondly, due to the limited computing capability, memory, or power supply of resource-constrained wireless devices, lightweight ML models are favored~\cite{Sharma2020toward}.
Thirdly, the distributed and dynamic nature of wireless networks restricts the available ML models and optimization techniques to a few options, such as GNNs~\cite{wu2021tnnls}, multi-agent reinforcement learning (MARL)~\cite{busoniu2008comprehensive}, and distributed optimization~\cite{giselsson2018large}. 
Although typically trained in a centralized and offline manner, GNNs can easily scale to large dynamic networks and generalize to different network topologies. 
This is because the same trained model of neural networks, which are often relatively lightweight, is shared by all nodes (or those of the same class).

Lastly, ML for wireless resource allocation also faces the challenge of dealing with various constraints. 
Soft constraints, which could be violated at a penalty, allow policies of distributed resource allocation to be directly generated by GNNs trained through primal-dual optimization \cite{wang2022learning} or imitation learning \cite{ross2011reduction}.
However, for combinatorial optimization (CO) formulated as mixed integer programming, ML systems must adopt certain algorithmic frameworks to ensure the hard constraints are always satisfied. 
Unfortunately, well-established sequential frameworks like Q-learning \cite{watkins1992q} and Monte-Carlo tree search \cite{silver2016mastering} are not applicable to distributed systems. 
A hybrid ML pipeline~\cite{zhao2021icassp,zhao2022twc,zhao2023iclr} can fill this gap by using a GNN to encode the network state into the inputs of a classical distributed algorithm, which outputs final decisions and
guarantees the observance of constraints on the decision variables.
Since many distributed algorithms are non-differentiable, training of GNNs in such hybrid ML pipelines used to rely on ZOO~\cite{liu2020primer}.
Recently, first-order deterministic policy gradient (DPG) \cite{silver2014deterministic} and graph-based DPG (GDPG) \cite{zhao2022twc,zhao2023iclr} have been applied to train GNNs for link scheduling and packet routing.
While our sparse scheduler is similar to GCN-LGS~\cite{zhao2022twc} in the use of non-differentiable operations and proxy function for gradient estimation, our learning algorithm takes one step further by simultaneously targeting two competing objectives: to minimize the signaling overhead while guaranteeing the total utility meets a minimum requirement.

\section{System Model and Problem Statement}
\label{sec:problem}

We model a wireless multi-hop network as a connectivity graph $\ccalG^n=(\ccalX,\ccalV)$, where each node $i\in\ccalX$ is a transceiver, and an edge $(i,j)\in\ccalV$ represents a directed link {from transmitter $i$ to receiver $j$ over-the-air (OTA).}
The sets of outgoing and incoming links of device $i$ are {$\kappa_i=\{{(i, j)}| {(i, j)}\in \ccalV \}$ and $\rho_i=\{ {(j, i)}| {(j, i)}\in \ccalV \} $.
Alternatively, edge $(i,j)\in\ccalV$ can also represent a bidirectional link between transceivers $i$ and $j$.} 
Our approach is applicable to both cases.
For each link, there is a queuing system $q$ for packet backlogs.

To describe the scheduling algorithm, we define the \emph{conflict graph}, $\mathcal{G}=(\ccalV,\ccalE)$, as follows: a vertex $v\in\ccalV$ represents a link in the wireless network, and the presence of an undirected edge $e=(v_a,v_b)\in\ccalE$ captures the interference relationship between links $v_a, v_b \in\ccalV$, which is considered to follow a physical distance model~\cite{cheng2009complexity}. 
For example, two links interfere with each other if their incident nodes are within a certain distance such that their simultaneous transmission would cause the outage probability to exceed a prescribed level, or they share the same node equipped with single radio interface.
For the rest of this paper, we assume the conflict graph $\mathcal{G}$ is known; see Section~\ref{sec:solution:conflict} for its estimation. 
We consider a simplified scenario in which all the nodes transmit at power levels that are time-invariant. 
In principle, the interference zone of each link (hence the entire $\mathcal{G}$) depends on the transmit power and directivity of the corresponding nodes.
Based on the definition of conflict graph $\mathcal{G}$, a legal schedule must be a set of wireless links that can be activated simultaneously subject to the constraint of {spatial OMA}, which forms an independent (vertex) set $\boldsymbol{v}$ on $\mathcal{G}$, defined as a set of nodes with no edges connecting each other.
{See~\cite[Sec.~III-C]{zhao2022twc} for multi-channel OMA extensions in time, frequency, and/or code dimensions}.

\begin{figure*}[t]
	\centering
	\hspace{-3mm}
	\subfloat[]{
		\includegraphics[width=0.26\linewidth]{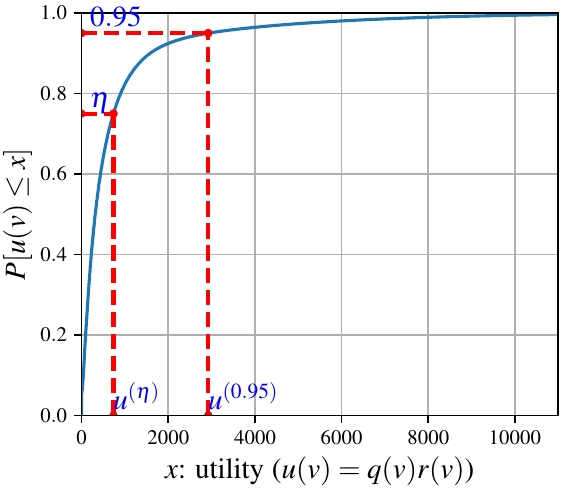}
		\label{fig:ecdf}
	 }%
	 \hspace{-3mm}
	\subfloat[]{
		\includegraphics[width=0.42\linewidth]{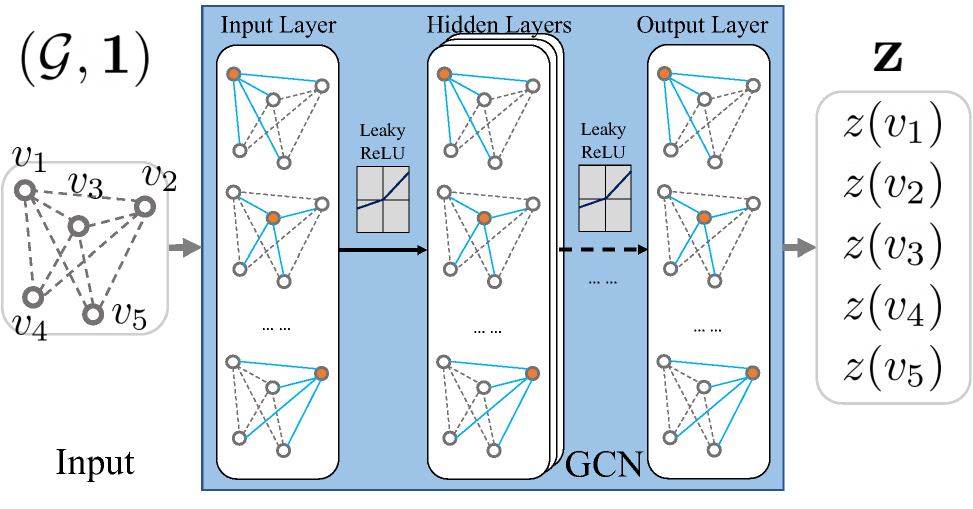}
		\label{fig:para}
	}  
	\subfloat[]{
		\includegraphics[width=0.28\linewidth]{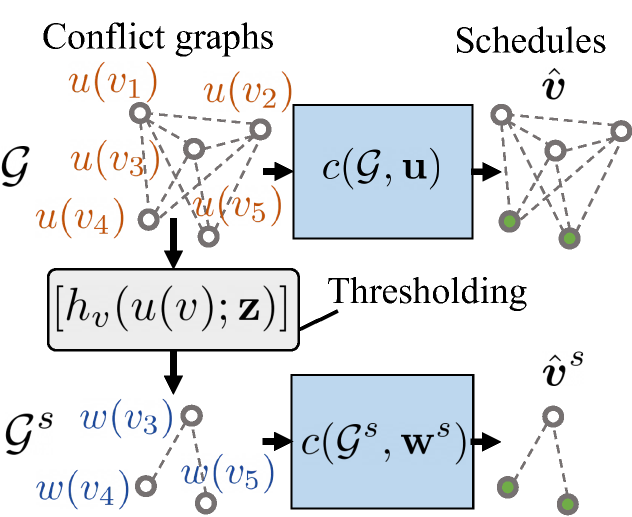}
		\label{fig:scheme}
	}  
	\caption{ Architecture of our GCN-based distributed sparse scheduler~\cite{zhao2022icassp_b}. (a) Select $u^{(\eta)}$ based on the eCDF of utility. (b) Parameters $\bbz$ generated by a featureless GCN based on the conflict graph $\ccalG$. (c) The sparse graph $\ccalG^s$ and topology-aware utility $\bbw^{s}$ created by the parameterized functions $ \{h_v(u(v); \bbz)\} $ based on $(\ccalG,\bbu)$. The distributed sparse scheduler generates schedule as $\hat{\boldsymbol{v}}^s=c(\ccalG^s,\bbw^s)$, whereas the dense scheduler generates schedules as $\hat{\boldsymbol{v}}=c(\ccalG,\bbu)$ for the purposes of training and comparison. 
	} \label{fig:system}
\end{figure*}

The \emph{network state} at time $t$, denoted as $(\mathcal{G}(t), \mathbf{u}(t))$, comprises the conflict graph $\mathcal{G}(t)$ that could change due to node mobility and a utility vector $\mathbf{u}(t)$ collecting $u(v,t)\in\reals_{+}$ for all nodes $v\in\ccalV$.
The utility $u(v,t)$ could capture the backlogs $q(v,t)$, link rate $r(v,t)$, sojourn time, and priority of the wireless link $v$. 
We denote by $c(\cdot)$ the scheduling contention process 
that maps every network state into an independent set on the conflict graph, {such as MaxWeight~\cite{joo2012local,joo2015distributed} and CSMA~\cite{ni2012qcsma,jiang2010distCSMA,Lin2009constant} schedulers.}
For notational simplicity, we henceforth omit $t$ for operations in the same time slot, and denote the total utility of a vertex set $\boldsymbol{v}$ by $u(\boldsymbol{v}) = \sum_{v\in\boldsymbol{v}}u(v)$.

Notice that in practice, global knowledge of network state is never required by any devices,
since the distributed executions of GNNs, link sparsification, and scheduling contention only involve interactions between each link and its conflicting neighbors.
Therefore, each link $v=(i,j)\in\ccalV$ only has to maintain its own utility $u(v,t)$ and the set of conflicting neighbors, denoted by $\ccalN_{\ccalG}(v)$.
The implementation of link operations on transceivers is detailed in Section~\ref{sec:solution:conflict}.

To minimize signaling overhead in the network, each link shall decide whether to contend for scheduling based on its own information, without talking with its neighbors.
Hence, we want to find vertex-specific functions $\{h_v(\cdot)\}_{v\in\ccalV}$ of local utility $u(v)$ to determine whether a vertex $v$ should join the contention.
Formally, we define our problem as follows.

\begin{Problem}\label{P:main} 
    Given a distribution $\Omega$ over network states $(\mathcal{G}, \mathbf{u})$ and a minimum required total utility $u_\mathrm{min}$, we want the optimal link sparsification functions $\{h_v^*\}$ for all $v \in \mathcal{V}$ as
    \begin{subequations}\label{eq:sp}
	\begin{align}
	 \{h_v^*\} &= \argmin_{\{h_v\}} \,\,\,  \mathbb{E}_{\Omega} \left(  |\ccalE^s | \right) \label{eq:sp:obj}\\[5pt]
	\text{s.t. } \,\,\, \ccalG^{s} & = \ccalG \setminus \{v| v\in\ccalV, h_v(u(v))\leq 0\}, \;\label{eq:sp:res}\\
    \ccalG^{s} &= (\ccalV^{s}, \ccalE^{s}),\;\label{eq:sp:gs} \\
	\bbw^s &= [h_v(u(v))], \; \forall\; v\in \ccalV^s,\; \label{eq:sp:ws}\\
	\hat{\boldsymbol{v}}^{s} &= c(\mathcal{G}^{s}, \mathbf{w}^s), \label{eq:sp:mwis}\\
	u_{\min} &\leq \mathbb{E}_{\Omega}\left[u(\hat{\boldsymbol{v}}^{s})\right]. \label{eq:sp:util}
	\end{align} 
    \end{subequations}
\end{Problem}

\vspace{1mm}
\noindent
To better understand Problem~\ref{P:main}, first notice that constraints~\eqref{eq:sp:res} and~\eqref{eq:sp:gs} define the sparsified conflict graph $\ccalG^s=(\ccalV^s,\ccalE^s)$ by removing from the original conflict graph $\ccalG$ those nodes (and, implicitly, all associated edges) with non-positive values of $h_v(u(v))$.
Thus, it is immediate that the form of functions $h_v$ has a direct influence on the level of sparsity of $\ccalG^s$.
Constraint~\eqref{eq:sp:ws} defines a modified utility vector $\bbw^s$ for the nodes of the sparsified graph $\ccalG^s$.
Furthermore, constraint~\eqref{eq:sp:mwis} determines the schedule  $\hat{\boldsymbol{v}}^{s}$ by applying a prescribed scheduler $c(\cdot)$ to $ (\ccalG^s, \bbw^s) $, and constraint~\eqref{eq:sp:util} specifies that the expected total utility of the schedule should be no less than a minimum value, which can be to set as the total utility of a schedule $\tilde{\boldsymbol{v}}^{s}$ under a baseline threshold policy, $u_\mathrm{min}=\mathbb{E}_{\Omega}\left[ u(\tilde{\boldsymbol{v}}^{s})\right]$, to take into account that the total utility is influenced by the size and density of networks. 
Our objective in~\eqref{eq:sp:obj} is to minimize the expected number of edges in our sparsified graph, which determines the message complexity of the contention process.
Finally, the inclusion of expectation in~\eqref{eq:sp:obj} says that we do not want to find sparsification functions for a particular network state but rather functions that generalize well across a whole distribution of network states.

Finding an exact solution to Problem~\ref{P:main} is extremely challenging for several reasons including:
i) The non-differentiable scheduler $c(\cdot)$ prevents direct application of gradient-based approaches,
ii) The optimization is over the space of functions, which is infinite-dimensional, and 
iii) The sparsification functions should be valid for a distribution of network states, possibly associated with conflict graphs of different sizes and topologies.
In the next section, we present our solution to Problem~\ref{P:main}, which addresses these challenges.

\section{Link Sparsification with GNNs}
\label{sec:solution}

In order to (approximately) solve Problem~\ref{P:main}, we 
consider a parametric family of thresholding functions $\{h_v\}$ given by~\eqref{E:param},
in which
 $z(v)=\bbz_v$ is the link parameter employed by link $v\in \ccalV$ to scale the prescribed global threshold,
$u^{(\eta)}$, which is the $\eta$-quantile utility under the network state distribution $\Omega$, as illustrated in Fig.~\ref{fig:ecdf}.
Notice that when $\bbz=\boldsymbol{1}$, the sparsification policy induced by $\{h_v\}$ in~\eqref{E:param} boils down to the baseline policy in~\eqref{E:baseline}.
Under the baseline, the Heaviside function compares the vertex utility $u(v)$ with $u^{(\eta)}$.
If $u(v)\leq u^{(\eta)}$, then $h_v(u(v)) = 0$ and the node is excluded from the sparse graph $\ccalG^s$ [cf.~\eqref{eq:sp:res}],
otherwise, $v$ is included in the sparse graph with the same utility $h_v(u(v)) = u(v)$.
In this context, our proposed parameterization in~\eqref{E:param} is a natural extension of the statistical baseline.

\subsection{Generation of the Link Parameters}

{To ensure that functions $\{h_v(\cdot;\bbz)\}$ can generalize to different underlying conflict topologies, we seek a function that can map a graph $\ccalG$ to the vector of link parameters $\bbz$.}
Specifically, 
we propose to have $\bbz = \Psi_{\ccalG}(\boldsymbol{1};\mathbf{\bbomega})$, 
where  $\Psi_{\ccalG}$ is an $L$-layered graph convolutional network (GCN) defined on the conflict graph $\ccalG$, and $\bbomega$ is the collection of trainable parameters of the GCN.
{In a GCN layer, each node updates its features by applying a non-linear activation to the sum of two linear projections: one of its own features and one of the aggregated features from its immediate neighbors. These projections are based on trainable parameters, enabling an $L$-layered GCN to learn how to combine information from each node and its $L$-hop local neighbors for a particular task.}

We define the output of an intermediate $l$th layer of the GCN as $\bbX^l \in\reals^{|\ccalV|\times g_{l}}$, and $\bbX^0 = \boldsymbol{1}^{|\ccalV|\times 1}$, $\bbz = \bbX^L $, then 
the $l$th layer of the GCN is expressed as:
\begin{equation}\label{E:gcn}
	\mathbf{X}^{l} = \sigma_l\left(\mathbf{X}^{l-1}{\bbTheta}_{0}^{l}+\bbcalL \mathbf{X}^{l-1}{\bbTheta}_{1}^{l}\right)\;, l\in\{1,\dots,L\}.
\end{equation}
In~\eqref{E:gcn},  
${\bbTheta}_{0}^{l}, {\bbTheta}_{1}^{l} \in \mathbb{R}^{g_{l-1} \times g_{l}}$ are trainable parameters, $\sigma_l(\cdot)$ is the element-wise activation function of the $l$th layer, and $ \bbcalL$ is the normalized Laplacian of $\ccalG$, i.e., $ \bbcalL = \bbI-\bbD^{-\nicefrac{1}{2}}\bbA\bbD^{-\nicefrac{1}{2}} $, where $\bbI$ is the identity matrix, $\bbA$ is the adjacency matrix of $\ccalG$, and $\bbD^{-\nicefrac{1}{2}}= \text{diag}(\left[\nicefrac{1}{\sqrt{d(v)}} \mid v\in\ccalV\right])$ where $d(v)$ is the degree of vertex $v$.
The activation functions of the input and hidden layers are selected as leaky ReLUs whereas a ReLU activation is used for the output layer. 
The input and output dimensions are set as $g_{0}=g_{L}=1$.

As a local operator on $\ccalG$, the normalized Laplacian $\bbcalL$ in \eqref{E:gcn} allows $z(v)$ to be computed in a distributed manner through neighborhood aggregation at $v$ with $L$ rounds of local exchanges between $v$ and its neighbors. 
Specifically, \eqref{E:gcn} can be implemented by the following local operation on link $v\in\ccalV$,
\begin{equation}\label{E:gcn:local}
    \bbX_{v*}^{l} \!=\! \sigma_l \!\left(\!\bbX_{v*}^{l-1} \, \bbTheta_{0}^{l} \!+ \!\left[\! \bbX_{v*}^{l-1} \!-\!\!\!\! \sum_{u \in \mathcal{N}_{\ccalG}(v)}\!\!\frac{\bbX_{u*}^{l-1}}{\sqrt{d({v})d({u})}} \!\right]\!\bbTheta_{1}^{l}\! \right)\!,
\end{equation}
where $\bbX_{v*}^{l}\in\reals^{1\times g_{l}}$ is the $v$th row of matrix $\bbX^{l}$, which captures the features on vertex $v$, $d(v)$ is the degree of vertex $v$, and $\ccalN_{\ccalG}(v)$ is the set of neighboring vertices of $v$ on $\ccalG$.

The downstream architecture of the entire distributed sparse scheduler~\cite{zhao2022icassp_b} is illustrated in Fig.~\ref{fig:system}.
First, the prior knowledge of eCDF of the utility values from empirical data of network operations is collected, as shown in Fig.~\ref{fig:ecdf}, based on which a global threshold is selected as $u^{(\eta)}$.
{In practice, this eCDF can be collected by a subset of devices based on the link utilities of themselves and their neighbors.}
Next, at the network level, the trained GCN observes the topology of the network and generates node embeddings as the link parameters $\bbz$.
Then, the sparse graph $\ccalG^s$ and topology-aware utility vector $\bbw^s$ are obtained by the parameterized functions $ \{h_v(u(v); \bbz)\} $  according to \eqref{eq:sp:res} and \eqref{eq:sp:ws}, respectively. 
Finally, the sparse schedule is obtained by the distributed scheduler as $\hat{\boldsymbol{v}}^s=c(\ccalG^s,\bbw^s)$ with lowered {scheduling} overhead.

We define local communication complexity as the rounds of local exchanges between a node and its neighbors.
The local communication complexity of a GCN is $\ccalO(L)$. 
Although each round of local exchange in GCN comprises $2|\ccalV||\ccalE|$ point-to-point messages, they can be implemented as $|\ccalX|$ broadcast transmissions in wireless networks (see Section~\ref{sec:solution:conflict}).
Hence, the local computational and communication costs can be controlled by modifying the number of layers $L$ in the GCN.
Importantly, this constant local communication complexity is critical to scalability.
Moreover, 
the link parameters $\bbz$ can be reused across time slots and only need to be updated when the topology changes, causing negligible impacts to wireless networks with relatively low mobility by diluting the overhead of generating $\bbz$ into hundreds to thousands of time slots.

\subsection{Transceiver-level Operations for Directed Links}\label{sec:solution:conflict}
Next, we discuss transceiver-level operations in the entire sparse scheduling pipeline, such as link maintenance (including conflict graph construction and hidden node problem avoidance), GCN operations, and scheduling contention.
We focus on the case of directed links as it is more relevant to massive access. 
The distributed operations of GCN in~\eqref{E:gcn:local} and contention $c(\cdot)$ only require a link $v$ to know its immediate conflicting neighbors $\ccalN_{\ccalG}(v)$ rather than the entire conflict graph $\ccalG$.
Therefore, the construction of $\ccalG$ is equivalent to each link $v\in\ccalV$ collecting and maintaining $\ccalN_{\ccalG}(v)$, which can be blended in the processes of initial access and recurrent contentions.
Conceptually~\cite[supplemental materials]{zhao2022twc}, we let the control messages of each link $v$ always carry its identity $v$, degree $d(v)$, and feature $\bbX_{v*}^{l}$, which enables a link to collect the IDs of its conflicting neighbors $\ccalN_{\ccalG}(v)$,  update $d(v)$, and implement GCN layers in~\eqref{E:gcn:local} by monitoring the channel over a sliding window of sufficient duration.

When a new device joins the network, it acquires a subchannel for sending control messages through initial access on a dedicated random access channel, e.g., by defective graph coloring on $\ccalG$ with a local complexity of $\ccalO(1)$ \cite{barenboim2013distributed}.
The transmitter of a link maintains its utility and performs sparsification based on its updated backlog information, whereas the receiver of link $v$ maintains its conflicting neighbors $\ccalN_{\ccalG}(v)$.
Each device $i\in\ccalX$ first decides a winner $v_i$ of internal contention among its undecided outgoing links, 
then it broadcasts a request-to-send (RTS) message, e.g., $\{i, d_i^o, \kappa_i, v_i, u(v_i)\}$, with its identity $i$, outgoing degree $d_i^o$ and outgoing link set $\kappa_i$, the identity and utility of link $v_i$.
Every device also listens to the control channel to collect RTS messages from nearby devices.
All the unique links detected by a device (with sufficient receive signal strength) over a sliding window, together with its own outgoing links, are in conflict with each other.
A device computes the GCN in~\eqref{E:gcn:local} and resolves OTA contentions for all of its incoming links. 
When a {directed link ${(i,j)}$} wins the OTA contention, its receiver $j$ broadcasts a clear-to-send (CTS) message $\{(i,j), d_j^{c}\}$ to notify the sender $i$ and mute its conflicting links $\ccalN_{\ccalG}((i,j))$, avoiding the hidden node problem.
Here, $d_j^{c}$ is the number of conflicting links sensed by device $j$.
A transmitter hearing this CTS learns that the conflict degrees of its outgoing links are at least $ d_j^{c}-1$.

The aforementioned transceiver-level principle can be translated into a distributed MaxWeight scheduler~\cite{joo2012local} in Algorithm~\ref{algo:lgs} and CSMA-based protocols~\cite{ni2012qcsma,jiang2010distCSMA,Lin2009constant}.
For CSMA, the contention is randomized and could be influenced by link utilities~\cite{ni2012qcsma}, in particular, the OTA contention for RTS messages is based on timing, where the first RTS picked up by a receiver wins.
When a device $j\in\ccalX$ finds that the first RTS it received is addressed to itself, it broadcasts a CTS message to notify the sender and mute other nearby transmitters.

\begin{algorithm}[t]
\caption{Local Greedy Solver (LGS) for Directed Links}
\label{algo:lgs}
\hspace*{\algorithmicindent} 
\textbf{Input}: $\ccalX,\ccalV,\ccalE, \ccalG^s=(\ccalV^s,\ccalE^s), \bbu\in\reals_{+}^{|\ccalV|}$ \\
\hspace*{\algorithmicindent} 
\textbf{Output}: $\bbv\!\in\!\{0,1\}^{|\ccalV|}$ \COMMENTS{0: mute, 1: scheduled}\\ \vspace{-0.15in}
\begin{algorithmic}[1] 
\STATE $\bbv=-\boldsymbol{1} + \mathbf{1}_{\{\ccalV\notin\ccalV^s\}}$, $\ccalG^n=(\ccalX,\ccalV),\ccalG=(\ccalV,\ccalE)$
\WHILE{ $\bbv_v\neq -1, \forall v\in\ccalV$}
\FORALL{ $i\in\ccalX$ }
\STATE $\ccalT_i=\{(i,j)| \bbv_{(i,j)}<0, j\in\ccalN_{\ccalG^n}(i)\}$, $d_i^{o}=|\ccalN_{\ccalG^n}(i)|$ 
\IF{ $ \ccalT_i\neq \emptyset $ }
\STATE $v_i=\argmax_{v\in \ccalT_i} \bbu_{v}$ \COMMENTS{Internal contention}
\STATE $\bbv_v=\mathbbm{1}(v=v_i),\;\forall v\in\ccalT_i$
\STATE Broadcast a RTS message $\left\{i, d_i^{o}, \kappa_i, v_i, \bbu_{v_i}\right\} $
\ENDIF
\STATE Receive RTS messages from nearby devices, denoted as set $\tilde{\ccalN}_{\ccalG}(i)\!=\!\big\{j \mid \exists x_1, x_2,\; ({(j,x_1)},\!{(x_2,i)})\!\in\!\ccalE \big\} $
\STATE $\Upsilon_i=\bigcup_{j\in\tilde{\ccalN}_{\ccalG}(i)\cup \{i\}}\{v_j\}$, $ d_i^{c} = \sum_{j\in\tilde{\ccalN}_{\ccalG}(i)\cup \{i\} }d_j^{o} $ 
\STATE $\tilde{v}_i=\argmax_{v\in \Upsilon_i} \bbu_{v}$ \COMMENTS{OTA contention}
\IF{$i$ is the receiver of link $ \tilde{v}_i $ }
\STATE $\bbv_{\tilde{v}_i}=1$, broadcast a CTS message with $(\tilde{v}_i, d_i^{c})$
\ELSIF{Receive CTS message $(\tilde{v}_j, d_j^{c})$} 
\STATE $\bbv_{\tilde{v}_j}=\mathbbm{1}(i\text{ is the transmitter of link } \tilde{v}_j)$
\ENDIF
\ENDFOR
\ENDWHILE
\end{algorithmic}
\end{algorithm}

\subsection{Asynchronous Implementation of GCNs}\label{sec:solution:async}
Since link parameters $\bbz$ are decoupled from the real-time link utilities $\bbu(t)$, an $L$-layer GCN can be computed asynchronously, e.g., when a new link joins the network, only its $L$-hop conflicting neighbors need to update their link parameters, for which the message exchanges can take many time slots by enclosing $\{\bbX_{v*}^{l}\}$ in the RTS and CTS for contentions.
In the distributed execution of a GCN layer, a device $i$ first exchanges a control message with its nearby devices, which encloses the features of their outgoing links, e.g., $\{\bbX_{v*}^{l-1}\}_{v\in\kappa_i}$.
From the received control messages, the device sorts out the features of its incoming links and their conflicting neighbors, then computes~\eqref{E:gcn:local} for each of its own incoming links, and exchanges the results $\{\bbX_{v*}^{l}\}_{v\in\rho_i}$ with nearby devices to obtain the output features of its outgoing links, $\{\bbX_{v*}^{l}\}_{v\in\kappa_i}$.

The most practical option is a 1-layer GCN with $L=1, g_{0}\!=\!g_{1}\!=\!1$, not only because of its effective yet lightweight model with just two trainable parameters, but also its friendliness to asynchronous implementation. 
Specifically, by enclosing $\{\bbX_{v*}^{l-1}\}_{v\in\kappa_i}$ in the RTS and $\{\bbX_{v*}^{l}\}_{v\in\rho_i}$ in the CTS for all devices $i\in\ccalX$,  
a newly joined device $j$ can learn the conflicting neighbors of its incoming links by passively monitoring the channel for a sufficiently long period, and get the link parameters $\{\bbz_{v}\}_{v\in\kappa_j}$ for its outgoing links with only one RTS-CTS cycle.
In networks with low mobility, such a lightweight, asynchronous 1-layer GCN can minimize the overhead of establishing the local threshold policies.

\subsection{Global Threshold}\label{sec:solution:global}
The optimal global threshold $u^{(\eta^*)}$ is typically set manually based on prior knowledge of the wireless networks. 
Ideally, the optimal cut-off quantile $\eta^*$ shall maximize the expected total utility net of signaling overhead, defined as follows
\begin{equation}\label{eq:eta}
\eta^* \!=\! \argmax_{0<\eta\leq 1}\; \mathbb{E}_{\Omega}\!\left\{\sum_{v\in {\bar{\boldsymbol{v}}(\ccalG^{s})}}\!\!\!\! u(v)\!\left[1 \!-\! f_{s}\!\bigl(d^{s}(v)\!+\!1,\tau\bigr) \right] \!\Bigg| \eta \!\right\} ,
\end{equation}
where $0<\tau\ll 1$ is the fraction of time slot (frame) for a link to broadcast a control message to its conflicting neighbors, $f_s(x,\tau)$ encodes that the scheduling overhead is a function of $\tau$ and the neighborhood size $ x $,
$d^{s}(v)$ is the expected degree of node $v$ on the sparsified graph $\ccalG^s$ if it is not cutoff, $\Omega$ is the distribution of network states $(\mathcal{G}, \mathbf{u})$, and $\bar{\boldsymbol{v}}(\ccalG^{s})$ is a maximal independent set on $\ccalG^s$ that contains the schedule, $\hat{\boldsymbol{v}}^s\subseteq \bar{\boldsymbol{v}}(\ccalG^{s})$.

For local greedy solver (LGS)-based protocols~\cite{joo2012local,zhao2021icassp,zhao2022twc}, the first round of message exchange dominates the entire point-to-point message complexity. 
If a sender started transmission upon receiving its CTS, denoted as \emph{flexible overhead}, the overhead can be approximated as $f_s(x, \tau)\!\approx\!\min(1, x\tau)$, otherwise, if all transmissions began only after a fixed scheduling deadline of $K\tau$, the overhead $f_s(x,\tau)\!\approx\!\mathbbm{1}( {x\tau > K\tau}) $, representing a link $v\in\bar{\boldsymbol{v}}(\ccalG^{s})$ is not scheduled due to signaling overdue.
In both cases, $f_s(x)$ monotonically increases with $x$.
For CSMA-based protocols with a fixed contention or backoff window $W$, the objective in~\eqref{eq:eta} can be rewritten as
$$ \mathbb{E}_{\Omega}\!\left\{\frac{1}{W}\sum_{v\in \ccalV^s}\!\! u(v)\sum_{m=0}^{W-1}\left(\frac{W-1-m}{W}\right)^{d^s(v)} \!\Bigg| \eta \!\right\}\;. $$

The approximated forms of $f_s(x)$ enable quick evaluation of the objective function in \eqref{eq:eta} for a given $\eta$ by sampling the graph $\ccalG\in\Omega$, the link utility $u(i)$, and the maximal independent set $\bar{\boldsymbol{v}}(\ccalG^s)$, which empirically exhibits quasi-concavity as shown in Fig.~\ref{fig:global:cutoff}. 
Therefore, the optimal $\eta^*$ can be found by an efficient search algorithm described in Appendix~\ref{apx:concave}.
In practice, $u^{(\eta)}$ can be refined periodically based on network dynamics.

\subsection{Hybrid Threshold Policy}\label{sec:solution:hybrid}
In practice, our GCN-based threshold policy can be conditionally applied to a link when the configuration of wireless protocols does not support its conflict degree:
\begin{equation}\label{eq:hybrid}
    h_v(u(v); \bbz) \!=\! u(v) \, H\!\left(u(v)- \mathbbm{1}(d(v) > D)z(v)u^{(\eta)}\right),   
\end{equation}
where $D$ is set to ensure that the overhead of the first round of local exchange $D\tau$ is within the scheduling sub-slot, e.g., $D<K$.
As shown in Section~\ref{sec:results:latency}, this hybrid policy can avoid unnecessary delays in less crowded parts of the network.

\section{First Order Optimization}\label{sec:foo}
Having discussed the downstream pipeline of the threshold-based link sparsification, we now turn to constrained unsupervised learning (UL) for training the GCNs.
Challenges stem from two aspects: 1)~the GCN is followed by non-differentiable operations, $\{h_v(\cdot)\}$ and $c(\cdot)$, and 2)~we have to balance two competing objectives: minimizing the overhead while achieving a minimum level of total utility.
To address these challenges, we reformulate Problem~\ref{P:main} as Problem~\ref{P:foo} {by introducing GCN-based parameterization of $\{h_v(\cdot)\}$, setting the minimum total utility to the statistical baseline, $u_\mathrm{min}=\mathbb{E}_{\Omega}\left[ u(\tilde{\boldsymbol{v}}^{s})\right]$, and bridging the non-differentiability with surrogate gradient in~\eqref{eq:obj:u}-\eqref{eq:grad:u}.}

\begin{Problem}\label{P:foo} 
    Given a distribution $\Omega$ over network states $(\mathcal{G}, \mathbf{u})$,
    where the utilities $u(v)\geq 0$ for $v\in\ccalV$ follow an i.i.d. distribution $\Omega_u$,
    we want to obtain the optimal set of parameters, $\bbomega$, for the GCN $\bbz=\Psi_{\ccalG}(\boldsymbol{1};\bbomega)$, where $\bbz_v\geq 0\; \forall\; v\in\ccalV$, as
    \begin{subequations}\label{eq:sp2}
	\begin{align}
	 \bbomega^* &= \argmin_{\bbomega\in\reals^{|\bbomega|}} \,\,\,  \mathbb{E}_{(\mathcal{G}, \mathbf{u})\sim\Omega} \Big( |\ccalE^s | \Big) \label{eq:sp2:obj}\\[3pt]
	\text{s.t. } \,\,\, \ccalG^{s} & = \ccalG \setminus \{v| v\in\ccalV, h_v(u(v);\bbz)\leq 0\}, \;\label{eq:sp2:res}\\
	\tilde{\ccalG}^{s} & = \ccalG \setminus \{v| v\in\ccalV, h_v(u(v); \boldsymbol{1})\leq 0\}, \;\label{eq:sp2:resb}\\
	\bbw^s &= [h_v(u(v); \bbz)], \; \forall\; v\in \ccalV^s,\; \label{eq:sp2:ws}\\
	\tilde{\bbw}^s &= [h_v(u(v); \boldsymbol{1})], \; \forall\; v\in \tilde{\ccalV}^s,\; \label{eq:sp2:wsb}\\
	\tilde{\boldsymbol{v}}^{s} &= c(\tilde{\mathcal{G}}^{s}, \tilde{\bbw}^s),\;\label{eq:sp2:vsb}\\
	\hat{\boldsymbol{v}}^{s} &= c(\mathcal{G}^{s}, \mathbf{w}^s), \label{eq:sp2:vs}\\
	0 &\leq \mathbb{E}_{\Omega}\left[u(\hat{\boldsymbol{v}}^{s}) - u(\tilde{\boldsymbol{v}}^{s})|(\ccalG,\bbu)\right], \;\label{eq:sp2:util}\\
	\bbz &= \Psi_{\ccalG}(\boldsymbol{1};\bbomega)\;. \label{eq:sp2:gcn}
	\end{align} 
\end{subequations}
\end{Problem}
The constraints in \eqref{eq:sp2:res}--\eqref{eq:sp2:vs} specify the respective processes of link sparsification under our approach and the statistical baseline. 
The constraint in \eqref{eq:sp2:util} sets the expected total utility of our approach to be no less than that of the statistical baseline.

\begin{algorithm}[t]
\caption{Alt-SGD: Alternating SGD for Constrained UL}
\label{algo:foo}
\hspace*{\algorithmicindent} \textbf{Input}: $\Omega,f_u(\cdot),\alpha, {0<\beta\leq 1}, N$, BatchSize \\
\hspace*{\algorithmicindent} \textbf{Output}: $\bbomega^*$ 
\begin{algorithmic}[1] 
\STATE Initialize $\bbomega$ randomly, $\ccalQ\leftarrow \varnothing$ 
\FOR{ epoch }
\STATE Draw $\eta\sim\left[0,1\right)$, $(\ccalG,\bbu)\sim\Omega$
\STATE Compute $\bbz$ based on \eqref{eq:sp2:gcn}, $u^{(\eta)}=f_u^{-1}(\eta)$.
\STATE Compute $u(\hat{\boldsymbol{v}}^{s}), u(\tilde{\boldsymbol{v}}^{s})$ using \eqref{eq:sp2:res}--\eqref{eq:sp2:vs}.
\IF{ $u(\hat{\boldsymbol{v}}^{s}) < u(\tilde{\boldsymbol{v}}^{s})$ }
\STATE $\widehat{\nabla_{\bbz}J(\bbomega)}=-\nabla_{\bbz} \mathbb{E}_{\Omega_{\bbu}^{\ccalG}} \left[ u(\hat{\boldsymbol{v}}^{s}) | \ccalG \right]$ based on \eqref{eq:grad:u}, \eqref{eq:proxy}  
\ELSE 
\STATE $\widehat{\nabla_{\bbz}J(\bbomega)}=\nabla_{\bbz}\mathbb{E}_{\Omega_{\bbu}^{\ccalG}} \left( |\ccalE^s | \big| \ccalG \right)$ based on \eqref{eq:grad:e}  
\ENDIF
\STATE $\widehat{\nabla_{\bbomega}J(\bbomega)} =\nabla_{\bbomega}\Psi_{\ccalG}(\boldsymbol{1},\bbomega)\widehat{\nabla_{\bbz}J(\bbomega)}$
\STATE $\widehat{\nabla_{\bbomega}J(\bbomega)}\leftarrow N \widehat{\nabla_{\bbomega}J(\bbomega)} / \big\|\widehat{\nabla_{\bbomega}J(\bbomega)} \big\|_2$
\STATE $\ccalQ \leftarrow \ccalQ \cup \{\widehat{\nabla_{\bbomega}J(\bbomega)} \}$
\IF{$ |\ccalQ|\geq $ BatchSize}
\FOR{$\widehat{\nabla_{\bbomega}J(\bbomega)} \in \ccalQ$} 
\STATE $\bbomega\leftarrow \bbomega - \alpha \widehat{\nabla_{\bbomega}J(\bbomega)}$
\ENDFOR
\STATE $\ccalQ\leftarrow\varnothing$, $\alpha \leftarrow \alpha \beta $
\ENDIF
\ENDFOR
\end{algorithmic}
\end{algorithm}

On an arbitrary graph $\ccalG$ with utility vector $\bbu\sim\Omega_{\bbu}^{\ccalG}$, the objective in \eqref{eq:sp2:obj} has an analytical form
\begin{equation}\label{eq:obj:e}
    \mathbb{E}_{\Omega_{\bbu}^{\ccalG}} \left( |\ccalE^s | \big| \ccalG \right) = \frac{1}{2} \sum_{v\in\ccalV}d^{s}(v)\left[1-p(v)\right]\;,
\end{equation}
where $d^{s}(v)$ can be computed as follows
\begin{equation}\label{eq:spdeg}
    d^{s}(v) = d(v) - \sum_{i\in\ccalN_{\ccalG}(v)}p(i).
\end{equation}
In \eqref{eq:spdeg},  $p(v)=f_u\left(z(v)u^{(\eta)}\right)$ is the cutoff probability of vertex $v$, where $f_{u}(\cdot)$ is the eCDF of per-link utility. 
Based on \eqref{eq:obj:e}, \eqref{eq:spdeg}, the probability density function (PDF) of link utility $f'_{u}(\cdot)$, {and the derivation in Appendix~\ref{apx:grad:e}}, we have 
\begin{equation}\label{eq:grad:e}
    \frac{\partial \mathbb{E}_{\Omega_{\bbu}^{\ccalG}} \left( |\ccalE^s | \big| \ccalG \right)}{\partial z(v)} =  -d^s(v)f_u'\left(z(v)u^{(\eta)}\right)u^{(\eta)}\;.
\end{equation}

We propose to solve~\eqref{eq:sp2} through alternating stochastic gradient descent (Alt-SGD), as described in Algorithm~\ref{algo:foo}, which minimizes the objective in~\eqref{eq:sp2:obj} while maintaining the inequality in~\eqref{eq:sp2:util}.
More precisely, when \eqref{eq:sp2:util} is satisfied, we update $\bbomega$ (in line 9) in the direction of the negative gradient $\nabla_{\bbomega}\mathbb{E}_{\Omega_{\bbu}^{\ccalG}} \left( |\ccalE^s | \big| \ccalG \right)=\nabla_{\bbomega}\Psi_{\ccalG}(\boldsymbol{1},\bbomega)\nabla_{\bbz}\mathbb{E}_{\Omega_{\bbu}^{\ccalG}} \left( |\ccalE^s | \big| \ccalG \right)$. 
Otherwise, we enforce satisfaction of~\eqref{eq:sp2:util} (in line 7) by updating $\bbomega$ in the direction of the gradient  $\nabla_{\bbomega}\mathbb{E}_{\Omega_{\bbu}^{\ccalG}} \left[u(\hat{\boldsymbol{v}}^{s}) \big| \ccalG\right] = \nabla_{\bbomega}\Psi_{\ccalG}(\boldsymbol{1},\bbomega)\nabla_{\bbz}\mathbb{E}_{\Omega_{\bbu}^{\ccalG}} \left[u(\hat{\boldsymbol{v}}^{s}) \big| \ccalG\right]$, to increase utility, where
\begin{equation}\label{eq:obj:u}
    \mathbb{E}_{\Omega_{\bbu}^{\ccalG}}\! \left[ u(\hat{\boldsymbol{v}}^{s})\big| \ccalG \right] \!=\! 
    \mathbb{E}_{\Omega_{\bbu}^{\ccalG}}\!\left[\bbu^{\top}\hat{\bbv}^{s} \big| \ccalG \right]\!=\!
    \sum_{v\in\ccalV}\mathbb{E}_{\Omega_{\bbu}^{\ccalG}}\!\left[\bbu_v\hat{\bbv}^{s}_{v}\big| \ccalG \right],
\end{equation}
and $\hat{\bbv}^{s}\in\{0,1\}^{|\ccalV|}$ is the indicator vector of set $\hat{\boldsymbol{v}}^{s}$ w.r.t. $\ccalV$.
Since $ \mathbb{E}_{\Omega_{\bbu}^{\ccalG}}\left[\bbu_v\hat{\bbv}^{s}_{v}\big| \ccalG \right] $ as a function of $p(v)$ does not have any analytical form, we consider two analytical proxies:
\begin{subequations}\label{eq:proxy}
\begin{align}
\mathbb{E}_{\Omega_{\bbu}^{\ccalG}}\left[\bbu_v\hat{\bbv}^{s}_{v}\big| \ccalG\right] &\approx a_1\left[1-p(v)\right]\;,\label{eq:proxy:1}\\
\mathbb{E}_{\Omega_{\bbu}^{\ccalG}}\left[\bbu_v\hat{\bbv}^{s}_{v}\big| \ccalG\right] &\approx a_2\left[1-p(v)\right]\left[1-a_3 d^{s}(v)\right]\;,\label{eq:proxy:2}
\end{align}
\end{subequations}
where $a_1, a_2>0$ and $0<a_3<1$ are constant parameters.
{The rationale behind \eqref{eq:proxy:1} is straightforward: the more likely a link $v\in\ccalV$ is retained after sparsification, the larger  is the expected utility gained from $v$.
The additional term $1-a_3d^s(v)$ in \eqref{eq:proxy:2} further captures the secondary factor that the expected utility gained on link $v$ is reduced if there are more conflicting neighbors of link $v$ after sparsification.}
Based on \eqref{eq:obj:u} and \eqref{eq:proxy}, we can find the gradient $\nabla_{\bbz} \mathbb{E}_{\Omega_{\bbu}^{\ccalG}} \left[ u(\hat{\boldsymbol{v}}^{s}) \big| \ccalG \right]$ as:
\begin{equation}\label{eq:grad:u}
    \frac{\partial \mathbb{E}_{\Omega_{\bbu}^{\ccalG}} \!\left[ u(\hat{\boldsymbol{v}}^{s}) \big| \ccalG \right]}{\partial z(v)} = \frac{\partial \mathbb{E}_{\Omega_{\bbu}^{\ccalG}}\!\left[\bbu_v\hat{\bbv}^{s}_{v}\big| \ccalG \right]}{\partial p(v)} f_u'\!\left(\!z(v)u^{(\eta)}\!\right)\!u^{(\eta)}.
\end{equation}
Since $f_u(\cdot)$ does not have an analytical form, we use a multilayer perceptron (MLP) $f_u(\cdot;\bbtheta)$ to fit $f_u(\cdot)$ in order to approximately compute the gradients in \eqref{eq:grad:e} and \eqref{eq:grad:u}.

Problem~\ref{P:foo} is non-convex due to the integer nature of the objective and constraints in \eqref{eq:sp2}.
Although the convergence of Algorithm~\ref{algo:foo} is not theoretically guaranteed, its intuition is as follows. 
The gradient in line 7 pushes the policy back into the constrained region of utility whenever it falls outside of it, whereas
the gradient in line 9 moves the policy towards lower cost.
With global norm being clipped in line 12, these two types of gradients share the same step size, but may not be in completely opposite directions. 
Therefore, the aggregation of sample gradients over a mini batch would likely point the policy to a direction of Pareto improvement, unless it is already in a locally Pareto optimal area within the constrained region, which is typically superior to the baseline policy.
By using a decaying learning rate (line 18), we gradually reduce the step size so that the policy can always converge after a prescribed number of epochs, which is our stop criterion.

\section{Numerical experiments}
\label{sec:results}

We evaluate the GCN-based threshold policy, $h_v(u(v);  \bbz)$, as a component of sparse schedulers in simulated time-slotted networks with synthetic random conflict graphs, which seek to represent wireless networks with uniformly distributed users of identical omnidirectional transmit power. 
Three GCNs with number of layers $L\in\{1,2,3\}$, denoted by GCN($L$), are evaluated, with an emphasis on GCN($1$).
{These policies are compared against} the statistical baseline with a global threshold, $h_v(u(v); \boldsymbol{1})$, as described in the beginning of Section~\ref{sec:solution}, and the zero-threshold policy, $h_v(u(v); \boldsymbol{0})$ (dense scheduler). 
These threshold policies are tested on four contention functions $c(\cdot)$, based on MaxWeight (LGS \cite{joo2012local}) and CSMA \cite{ni2012qcsma}.

\subsection{Experimental Configurations}
\label{sec:results:config}

\begin{figure*}[t!]
	\centering
	\subfloat[]{
		\includegraphics[height=1.4in]{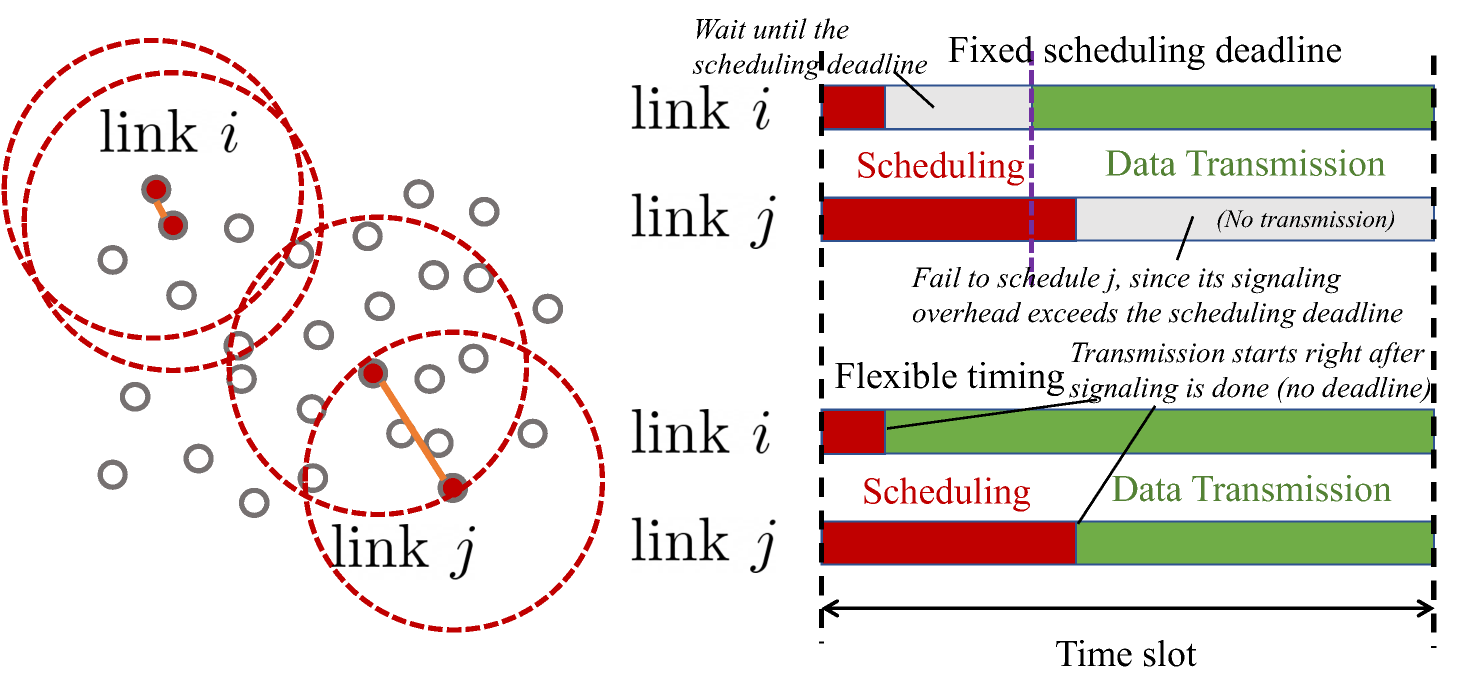}
        \label{fig:timing}  
		\vspace{-0.05in}
	}\hspace{-4mm}
    \subfloat[]{
        \includegraphics[height=1.4in]{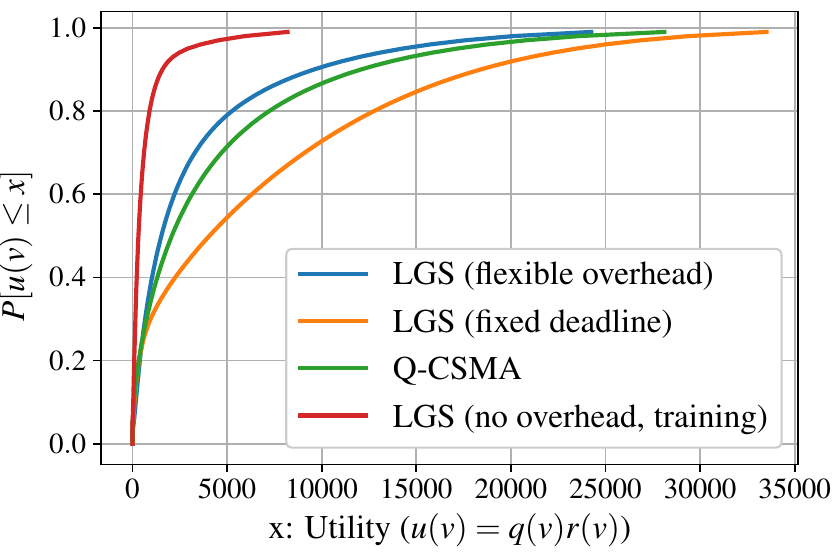}
        \label{fig:util:ecdf}  
        \vspace{-0.05in}
    }\hspace{-4mm}
	\subfloat[]{
		\includegraphics[height=1.4in]{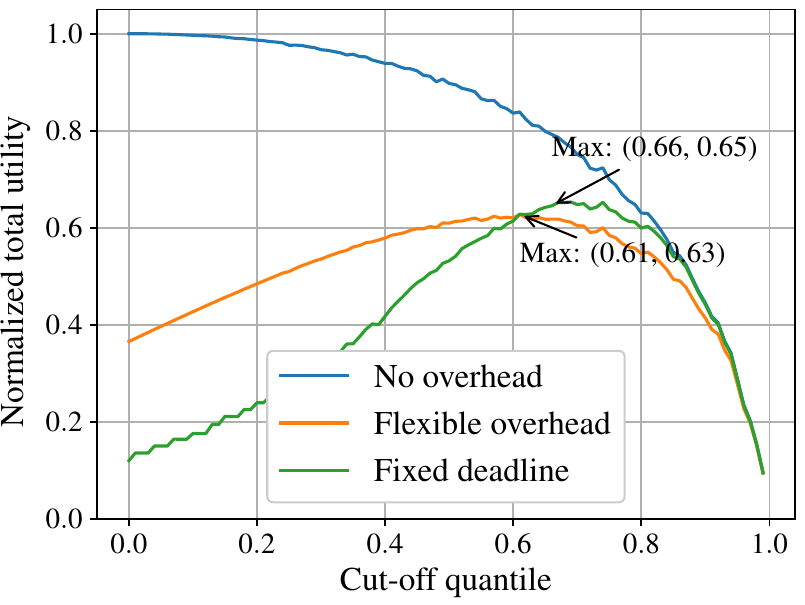}
    \label{fig:global:cutoff}
	 	\vspace{-0.05in}
	}
	\caption{a) Illustration of two LGS-based protocols: With fixed scheduling deadline, a link will be scheduled only if the signaling meets the deadline, and transmissions start after the deadline; with flexible timing, data transmissions start immediately after the signaling, when all the interfering neighbors of the scheduled links are muted~\cite{zhao2022twc}. 
    b) The global distribution (eCDFs) of per-link utility collected from simulation on ER test set under four distributed scheduling protocols: ideal LGS with no overhead, two LGS-based protocols, and Q-CSMA. 
    c) Normalized total utility net of signaling overhead (AR) vs cut-off quantile of global threshold under the two LGS-based scheduling protocols.
	}   
	\label{fig:global}    
\end{figure*}

The training settings include a learning rate of $\alpha=10^{-4}$, $\beta=0.996$, $N=0.05$, a batch size of 100, and a total of 25 epochs.\footnote{Training takes {10-20} hours on a workstation with a specification of 16GB memory, 8 cores, and Geforce GTX 1070 GPU. The source code is published at \url{https://github.com/zhongyuanzhao/gcn-sparsify}}
The network states $(\ccalG, \bbu)$ for training are generated as follows:
The link utility $u(v)\!=\!\bbu_v, v\in\ccalV$ is independently drawn from an empirical distribution (see Figs.~\ref{fig:ecdf} and~\ref{fig:util:ecdf}), which is collected from simulation of a specific scheduling protocol on a set of synthetic conflict graphs and the following network traffic configurations.

\textbf{Network traffic:} The link utility function is $u(v,t)=q(v,t)r(v,t)$ \cite{joo2012local}, and the backlog (queue length) evolves as $$q(v,t\!+\!1)\!=\! q(v,t)\!-\!\bbv_v\!(t)\min\{r(v,t),q(v,t)\} \!+\! a(v,t),$$ where $a(v,t)$ is the number of exogenous packets arriving at the source of link $v$ at time step $t$, $ \bbv_v\!(t)=1 $ (or $ 0 $) if  link $v$ is (not) scheduled at time $t$. 
The nominal link rate $r(v,t)$ for each link and time step is independently drawn from a normal distribution $\mathbb{N}(50, 25)$, rounded up, and then clipped to $\left[0,100\right]$, 
which captures the effects of fading and lognormal shadowing in wireless channel\cite{Mousavi17lte}.
For simplicity, the data traffic is configured as a single hop flow generated for each link $v\in\ccalV$. 
For each flow, the exogenous packets at the source user follow a Poisson arrival with a prescribed arrival rate $\lambda$.
The traffic load, defined as {$\mu = \lambda/\mathbb{E}_{v\in\ccalV, t\leq T}\left[r(v,t)\right]$}, 
is set as $\mu\in\left[0.03,0.05\right]$ for all threshold policies,
offering unsaturated traffic for sparser wireless networks, i.e., the average conflict degree $\bar{d}\leq25$.
Each scheduling instance includes a conflict graph $\ccalG=(\ccalV,\ccalE)$, 
and realizations of random packet arrivals $a(v,t)$ and link rates $r(v,t)$ for $v\in\ccalV, 0< t\leq T$ where $T=200$.
Although each scheduling instance has a static topology, we test a trained GCN on many conflict graphs with different sizes and topologies without any re-training, therefore, the GCN-based threshold policies can generalize to dynamic topology due to network mobility.

\begin{figure}[t!]
	\centering
	\subfloat[]{
		\includegraphics[height=1.4in]{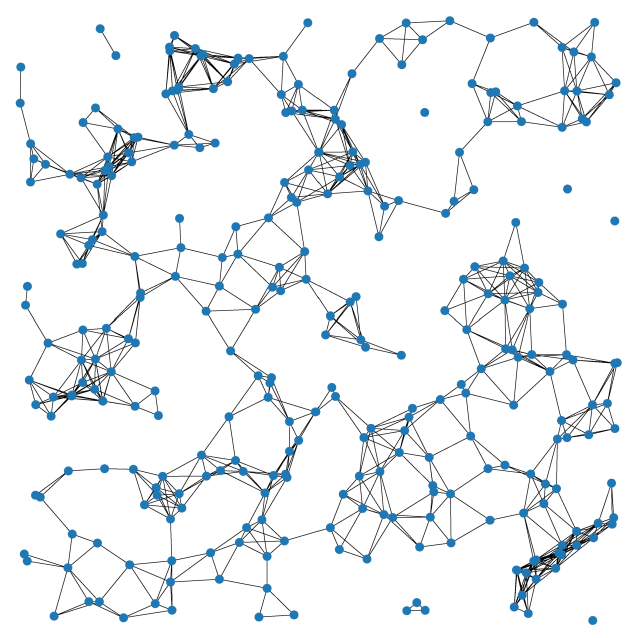}
		\label{fig:justify:example}
	}
	\subfloat[]{
		\includegraphics[height=1.33in]{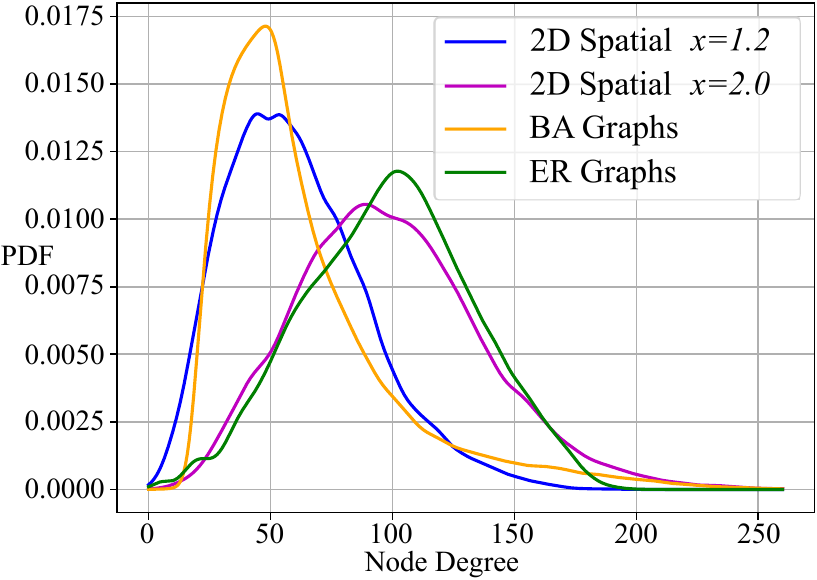}
		\label{fig:justify:degree}
	}
	\caption{Random instances of wireless multi-hop networks generated from a 2D spatial process, with two ratios of interference to connectivity radii, $x\in\{1.2, 2.0\}$:
    a) An instance of wireless networks, where an edge represents a link.
    b) The degree distributions of 100 instances of conflict graphs with $x=1.2$ and $x=2.0$,  the BA test set in Table~\ref{tab:graphs}, and another set of ER graphs (not listed in Table~\ref{tab:graphs}) with $|\ccalV|\in \lfloor\mathbb{U} (200, 400)\rceil$ and $\bar{d}\in\mathbb{N}(100,30)$.
	}   
	\label{fig:justify}    
\end{figure}

\begin{table}[t!]
	\renewcommand{\arraystretch}{1.15}
	\caption{Configurations for datasets of synthetic random graphs} 
	\label{tab:graphs}\vspace{-0.05in}
	\centering
    \setlength{\tabcolsep}{4pt}
	\footnotesize
	\begin{tabular}{|p{1.4cm}|p{4cm}|p{1.6cm}|c|}
    \hline
	\textbf{Set (size)}  & \textbf{Parameters} & \textbf{Quantity} & \textbf{Total}  \\ \hline
    \multirow{2}{1cm}{ER training set (5900)} & $|\ccalV|\in \left\{100, 150, 200, 250, 300\right\}$,\newline $\bar{d}=|\ccalV|k\in \left\{2, 5, 7.5, 10, 12.5\right\}$ & 200 instances \newline per $(|\ccalV|, \bar{d})$ & 5000 \\ \cline{2-4}
    & $|\ccalV|\in\left\{30,100\right\}$, \newline $k\in\left\{0.1,0.2,\dots,0.9\right\}$ & $50$ instances \newline per $(|\ccalV|, k)$ & 900 \\ \hline
    \multirow{2}{1cm}{BA training set (5900)} & $|\ccalV|\in \left\{100, 150, 200, 250, 300\right\}$,\newline $m\in \left\{2, 5, 7.5, 10, 12.5\right\}$ & 200 instances \newline per $(|\ccalV|, m)$ & 5000 \\ \cline{2-4}
    & $|\ccalV|\in\left\{30,100\right\}$, $m=|\ccalV|k$, \newline $k\in\left\{0.1,0.2,\dots,0.9\right\}$  & $50$ instances \newline per $(|\ccalV|, k)$ & 900 \\ \hline
    ER test set (500) & $|\ccalV|\in \left\{100, 150, 200, 250, 300\right\}$,\newline $\bar{d}=|\ccalV|k\in \left\{2, 5, 10, 15, 20\right\}$ & 20 instances \newline per $(|\ccalV|, \bar{d})$ & 500 \\ \hline
    \multirow{2}{1cm}{BA test set (860)} & $|\ccalV|\in \left\{100, 150, 200, 250, 300\right\}$,\newline $m\in \left\{2, 5, 10, 15, 20\right\}$ & 20 instances \newline per $(|\ccalV|, m)$ & 500 \\ \cline{2-4}
    & $|\ccalV|\in\left\{300,400,500\right\}$, \newline $m\in\left\{25,30,35,40,45,50\right\}$  & $20$ instances \newline per $(|\ccalV|, m)$ & 360 \\ \hline
	\end{tabular}
    \begin{tabular}{|l|l|c|c|}
    \hline
	\textbf{Section}  & \textbf{Experiments} & \textbf{Training set} & \textbf{Test set}  \\ \hline
    \ref{sec:results:graphs} & Total utility with identical inputs & ER-train & ER-test \\ \hline
    \ref{sec:results:deadline} & Latency with LGS (fixed deadline) & ER-train  & BA-test \\ \hline
    \ref{sec:results:overhead} & Latency with LGS (flexible overhead) & BA-train  & BA-test \\ \hline
    \ref{sec:results:csma} & Latency with two CSMA schedulers & ER-train  & BA-test \\ \hline
	\end{tabular}
 \vspace{-0.1in}
\end{table}

\begin{figure*}[t]
	\centering
	\subfloat[]{
		\includegraphics[width=0.45\linewidth]{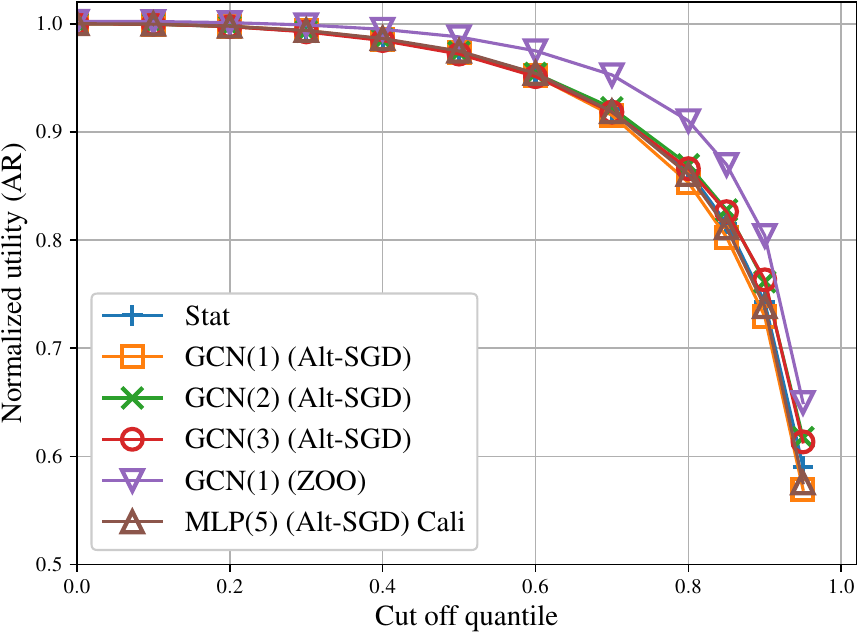}
		\label{fig:results:util} 
		\vspace{-0.05in}
	}
	\subfloat[]{
		\includegraphics[width=0.45\linewidth]{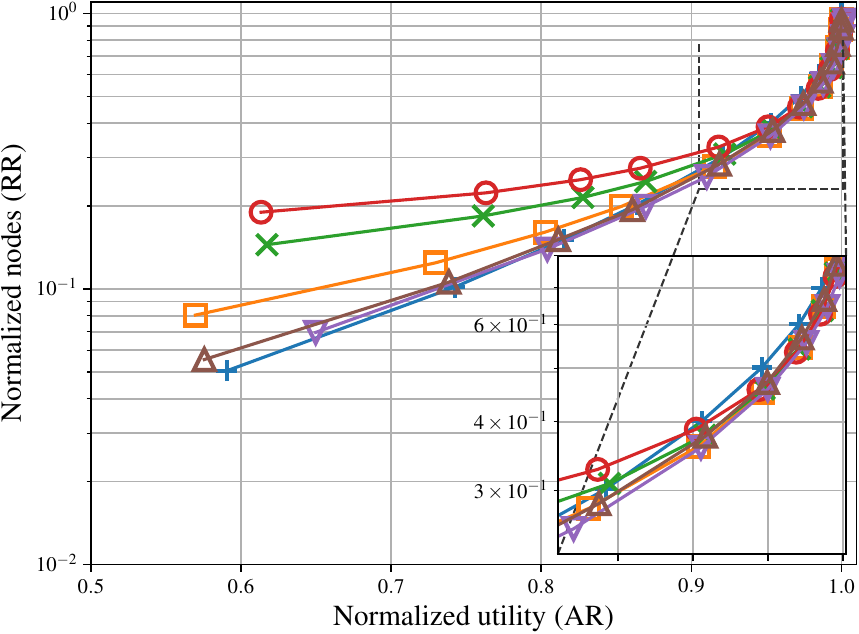}
		\label{fig:results:size}
		  \vspace{-0.05in}
	}\\ 
	\subfloat[]{
		\includegraphics[width=0.45\linewidth]{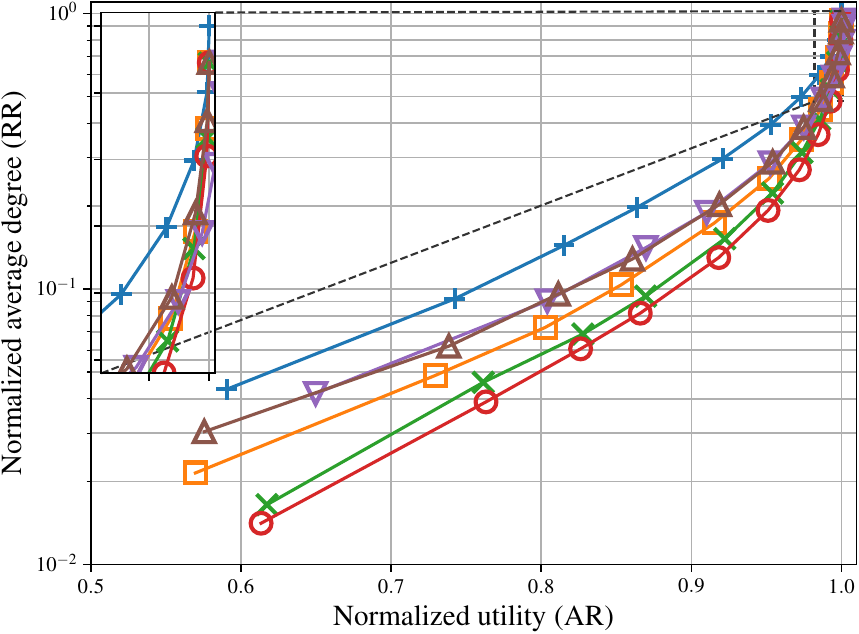}
		\label{fig:results:deg}
		\vspace{-0.05in}
	}
	\subfloat[]{
		\includegraphics[width=0.45\linewidth]{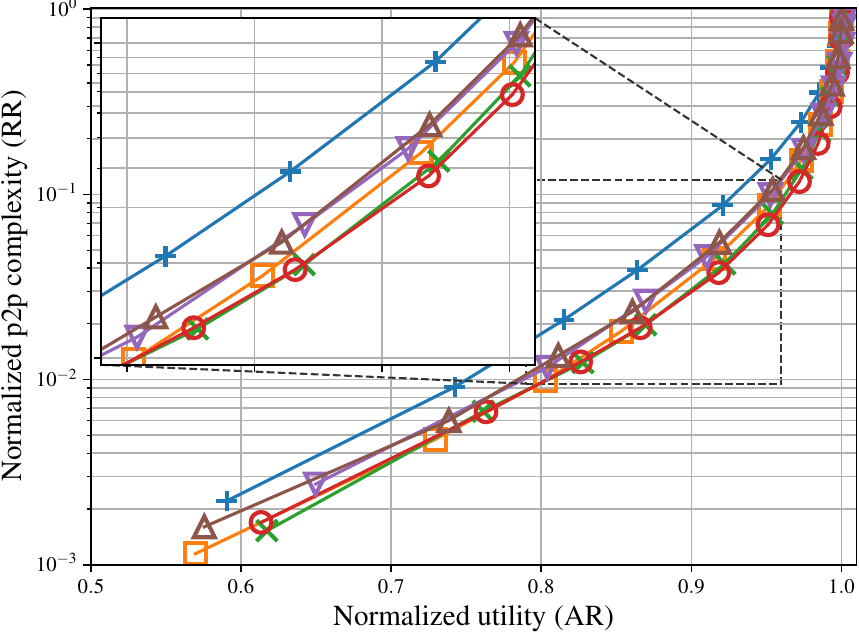}
		\label{fig:results:p2p}
		\vspace{-0.05in}
	}
	\caption{The performance of different threshold policies normalized by that of zero-threshold policy with identical input network state $(\ccalG,\bbu)$, under vanilla LGS \cite{joo2012local}:  
    (a) AR for total utility vs cut-off quantile $\eta$ showing that Alt-SGD can align GCN-based and the baseline threshold policies on total utilities, (b) RR for number of vertices in $\ccalG^s$ vs normalized utility, (c) RR for average degree of $\ccalG^s$ vs normalized utility, and (d) RR for P2P message complexity of scheduling contention vs normalized utility. 
		For (c) and (d), smaller RR is better.
	}   
	\label{fig:results}    
\end{figure*}

\textbf{Conflict graph:}
Four datesets of synthetic conflict graph $\ccalG$ for training and testing are generated from the Erdős–Rényi (ER)~\cite{erdds1959random} model and Barabási–Albert (BA)~\cite[Ch~5]{barabasi2016network} model, as detailed in Table~\ref{tab:graphs}, where $|\ccalV|$ is the size of graph, $k$ is the probability of edge-appearance in ER model, $m$ is the preferential attachment parameter in BA model, and $\bar{d}$ is the average degree of a graph.
For BA graphs, $\bar{d}=2m$ \cite[Ch~5]{barabasi2016network}.
For each experiment, the training and test sets may or may not be from the same graph model, as shown in Table~\ref{tab:graphs}.
Notice the slight mismatch between training and test sets, i.e., $\bar{d} = 15,20$ ($m\geq15$, $|\ccalV|>300$) are not in the ER (BA) training set. 

To illustrate how the ER and BA models can capture the conflict graphs, we compare the degree distributions of the ER and BA test sets with the conflict graphs of 100 instances of wireless multi-hop networks. 
Each network instance is generated from a 2D spatial process, in which 300 wireless devices, equipped with uniform transmit power and omnidirectional antennas, are randomly located in a $12\times 12 $ square; a link (edge) is formed if two devices (nodes) are within a distance of $1.0$; and two links interfere with each other if any of their incidental nodes are within a distance of $x \in \{1.2,2.0\}$. 
{For generality, the unit of distance is omitted  since path loss is a function of both carrier frequency and distance.}
The nodes and links in one of these instances are visualized in Fig.~\ref{fig:justify:example}. 
The PDFs of conflict degree of these network instances are presented in Fig.~\ref{fig:justify:degree}.
With smaller interference radius ($x=1.2$), the distribution of node degrees of the conflict graphs is right-skewed, matching the degree distribution of graphs in the BA test set.
With larger interference radius ($x=2.0$), the conflict degree distribution is more symmetric and similar to a set of ER graphs (not in Table~\ref{tab:graphs}) with $|\ccalV|\in \lfloor\mathbb{U} (200, 400)\rceil$ and $\bar{d}\in\mathbb{N}(100,30)$.

\subsection{Overhead Under Identical Network State}
\label{sec:results:graphs}

We first compare the overheads of various threshold policies under LGS schedulers per time-slot, given identical network states and cut-off quantiles. 
For an input network state $(\ccalG(i), \bbu(i))$, the conflict graph $\ccalG(i)$ is drawn from the ER test set, and the link utility $\bbu_{v}(i), v\in\ccalV$ is drawn from the empirical distribution of per-link utility (Fig.~\ref{fig:ecdf}) collected from an ideal LGS scheduler with realized link rate equals to the nominal one, under the network traffic setting and ER test set described in Section~\ref{sec:results:config}.
Each network state is tested with 12 cut-off quantiles $ \eta\in\{0, 0.1,\ldots, 0.8, 0.85, 0.9, 0.95\} $.

Besides the regular GCN($L$) with $L\in\{1,2,3\}$ trained with Alt-SGD, we also train a GCN(1) with ZOO to serve as the baseline for Alt-SGD. 
Additionally, to evaluate the benefit of selecting GCN as the ML model, we train another 5-layer perceptron with Alt-SGD, denoted by MLP(5), which resides on each node and takes the degree of its own node as the only input -- without the neighborhood aggregation in GNNs. 
Notice that this MLP(5) is different from the MLP $f_u(\cdot;\bbtheta)$ used in \eqref{eq:grad:e} and \eqref{eq:grad:u} for training.

In Fig.~\ref{fig:results}, the performance of the tested threshold policies is presented in relative formats, as approximation ratio (AR) or retention ratio (RR), of which the nominator is a metric of the tested threshold policy and the denominator is that of the zero-threshold policy.
In Fig.~\ref{fig:results:util}, the ARs of the total utility achieved by GCNs and MLP(5) trained by Alt-SGD are well aligned with the statistical baseline (Stat) across all cut-off quantiles $\eta$, whereas the GCN(1) trained by ZOO always achieve larger utility at a higher cost for given $\eta$, with a cost-efficiency similar to MLP(5) as shown next.
It demonstrates that Alt-SGD is more effective than ZOO in balancing two competing goals, i.e., minimizing the overhead while achieving required levels of total utility, whereas GCN(1) trained by ZOO would require an additional step of calibration.

To get a deeper insight into the behaviors and cost efficiencies of the tested threshold policies, we plot their RRs of the size of the sparsified conflict graph in~Fig.~\ref{fig:results:size}, the average post-sparsification conflict degree (for $\ccalG^s$), which is linear in the collision rate in CSMA in Fig.~\ref{fig:results:deg}, and their RRs of the point-to-point (P2P) message complexity of LGS in Fig.~\ref{fig:results:p2p}, all against the ARs of total utility (normalized utility).
The slope of a point in Fig.~\ref{fig:results:deg} (Fig.~\ref{fig:results:p2p}) represents the overhead cost of achieving an unit of utility under the CSMA (LGS) protocol.
Therefore, the lower a curve lies in Figs.~\ref{fig:results:deg} and~\ref{fig:results:p2p}, the more efficient is the corresponding policy.
For both CSMA and LGS protocols, all the ML-based threshold policies are more cost-effective than the statistical baseline.
Among the ML-based policies, the three GCNs trained by Alt-SGD are more cost-effective than MLP(5) trained by Alt-SGD and GCN(1) trained by ZOO, where the latter two have similar signaling efficiencies. 
It shows that both the GNN model and Alt-SGD algorithm made meaningful contributions to the performance.
Moreover, Alt-SGD is able to exploit the expressiveness of deeper GCNs in learning topological information, thus empowering them to achieve greater efficiencies than the shallower ones,
in the region of deep sparsification, e.g., $\eta\geq 0.7$ or normalized utility $\leq 0.95$.
Fig.~\ref{fig:results:size} further shows that under the same normalized utility, deeper GCNs are better at picking links, i.e., retain more links for contention while achieving lower scheduling overhead.

\subsection{Latency in Scheduling Experiments}
\label{sec:results:latency}

In link scheduling, identical inputs $\bbu(t)$ for different sparse schedulers rarely happen because of the dependency between $\bbu(t)$ and scheduling decisions.
To illustrate how the improved signaling efficiency in Section~\ref{sec:results:graphs} could be translated to better network performance, we compare six threshold policies with the zero-threshold policy (Dense-Zero) in simulated wireless multi-hop networks under four different distributed scheduling protocols.

The tested threshold policies include 1) Sparse-Stat: global threshold, $u^{(\eta)}$, as the statistical baseline with a given $\eta$; 2) Sparse-Stat-(1): global threshold  $\mathbb{E}_{v\in\ccalV}({\bbz_v})u^{(\eta)}$ that takes into account the scaling effect of the 1-layer GCN;
3) Sparse-GCN($L$): the GCN-based local threshold policies, with the GCNs trained by Alt-SGD in Section~\ref{sec:results:graphs}; and
4) Hybrid-GCN(1): the hybrid threshold policy described in \eqref{eq:hybrid}, which enables the GCN(1)-based local threshold on a link $v$ only if its conflict degree {$d(v)>D=25$}, so that $D\tau < 30\tau$, the expected scheduling deadline defined later.
Hybrid-GCN(1) can demonstrate the performance of our threshold policies in practical LGS-based MaxWeight scheduling.

The first two distributed scheduling protocols are LGS-based MaxWeight schedulers with different ways of splitting a time slot into scheduling and communication sub-slots, as illustrated in Fig.~\ref{fig:timing}~\cite{zhao2022twc}.
In \textit{LGS with fixed deadline}, scheduling stops by a prescribed deadline, e.g., at 30 ms of a time slot of 100 ms, and only the links that are successfully scheduled by the deadline can use the rest of the time slot, e.g., 70 ms, for transmissions.  
In \textit{LGS with flexible overhead}, transmission starts right after a link is scheduled and broadcasts a control message to mute its interfering neighbors. 
Since the interfering neighbors of scheduled links are all muted, such transmission can cause no disruptions. 
We consider that it takes 1 ms for a link to broadcast a control message to its interfering neighbors, i.e., a full round of message exchange in a neighborhood of 20 links would take 20 ms.

The eCDFs of link utility under zero-threshold policy with ideal LGS scheduler, LGS with fixed deadline and flexible overhead, and Q-CSMA are illustrated in Fig.~\ref{fig:util:ecdf}, where for the latter three, the realized link rate differs from the nominal one.
We train two sets of GCNs based on 1) the ER training set and the utility eCDF for LGS with fixed deadline, and 2) the BA training set and the eCDF for LGS with flexible overhead.
The first set of GCNs are then reused in another two scheduling protocols, unweighted CSMA and Q-CSMA in Section~\ref{sec:results:csma}, to evaluate their generalizability.

Based on Section~\ref{sec:solution:global}, the optimal cutoff quantiles for the BA test set under LGS with fixed deadline and flexible overhead are estimated 
 as $\eta=0.66$ and $0.61$, respectively, as shown in Fig.~\ref{fig:global:cutoff}.
Notice that there is a deliberate mismatch in interference density $\bar{d}$ between the entire BA test set and our target range $60\leq \bar{d}\leq 100$ to reflect reality.

\begin{figure}[t!]
	\centering
	\subfloat[]{
		\includegraphics[width=0.9\linewidth]{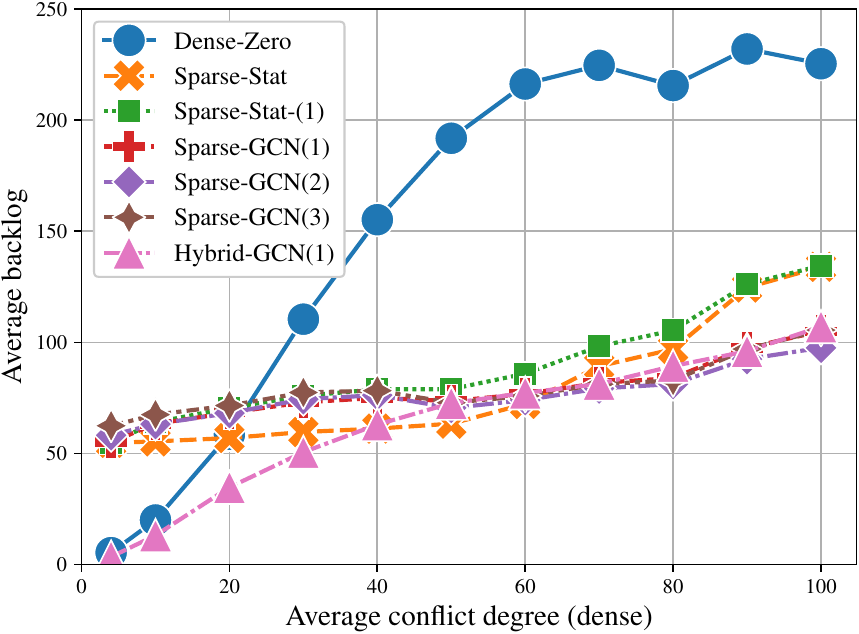}
		\label{fig:deadline:backlog}
	}\\
	\subfloat[]{
		\includegraphics[width=0.9\linewidth]{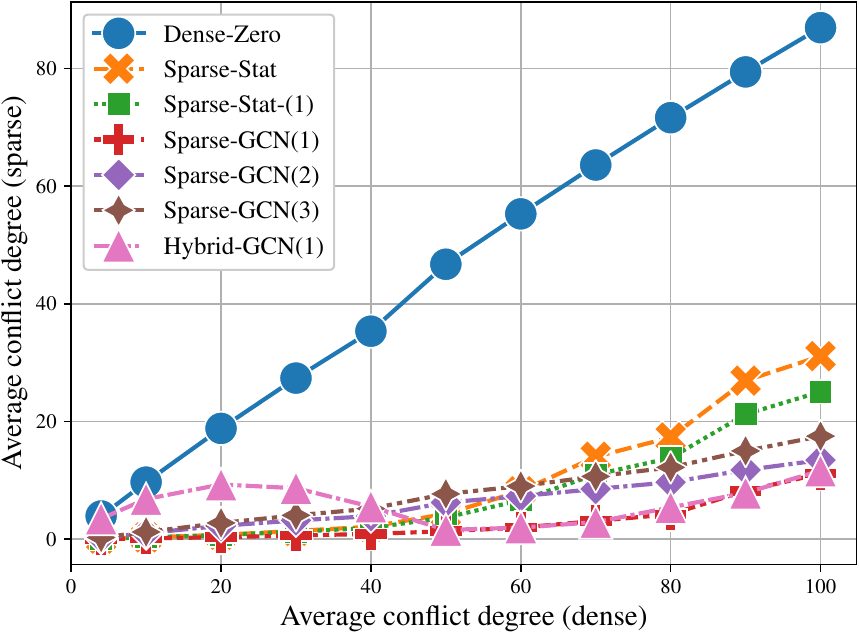}
		\label{fig:deadline:degree}
	}
	\caption{The performance of threshold policies vs average conflict degree $\bar{d}$ under LGS with fixed deadline:
	(a) average backlog, and
    (b) average post-sparsification conflict degree,  averaged across time and network. 
	Each point is the average of 60-100 test instances. Global cut-off quantile $\eta=0.66$.
    GCNs are trained on ER training set.
	}   
	\label{fig:deadline}    
\end{figure}

\begin{figure}[t]
	\centering
	\subfloat[]{
		\includegraphics[width=0.9\linewidth]{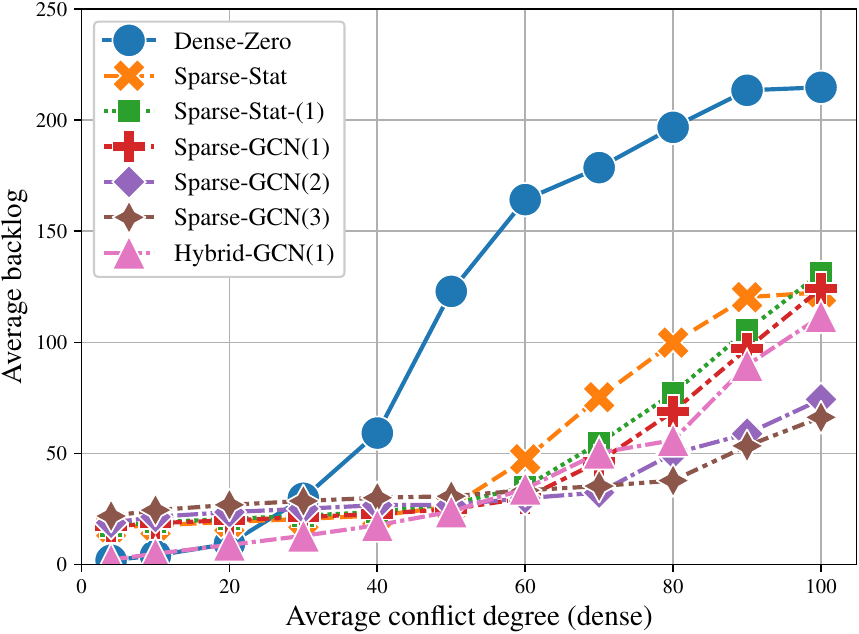}
		\label{fig:overhead:backlog}
	}\\
	\subfloat[]{
		\includegraphics[width=0.9\linewidth]{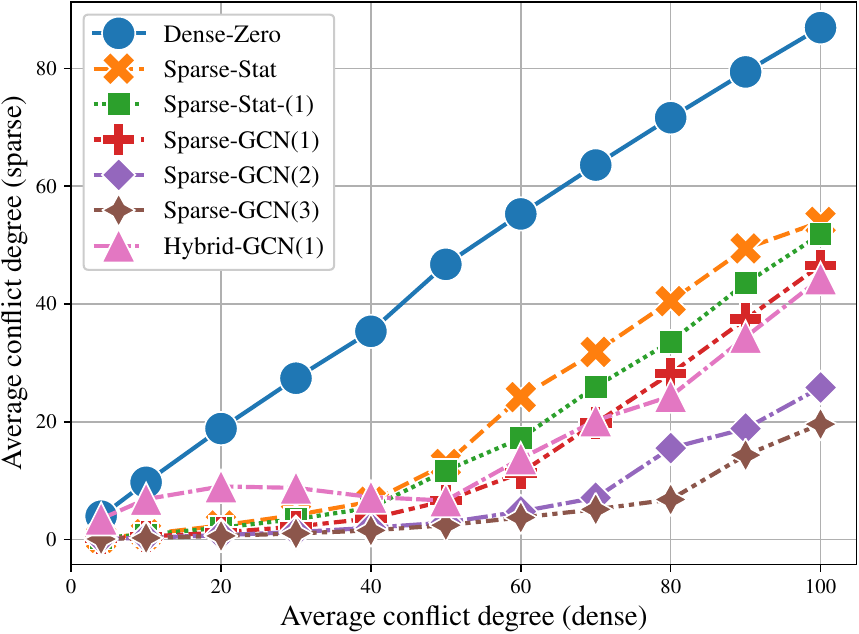}
		\label{fig:overhead:degree}
	}
	\caption{The performance of threshold policies vs average conflict degree $\bar{d}$ under LGS with flexible overhead:  
	(a) average backlog, and
    (b) average post-sparsification conflict degree, averaged across time and network. 
	Each point is the average of 60-100 test instances.
    $\eta=0.61$.
    GCNs are trained on BA training set.
	} 
	\label{fig:overhead}    
\end{figure}

\subsubsection{LGS with Fixed Deadline}\label{sec:results:deadline}

In this experiment, the realized link rate is $r(v,t)$ if a link $v$ is scheduled by the deadline (otherwise $0$), the traffic setting and BA test set described in Section~\ref{sec:results:config} are utilized, with a global threshold $u^{(\eta)}=7912$ for $\eta=0.66$.
The latency performances of tested threshold policies, measured by average backlog across time and networks, as a function of the average conflict degree $\bar{d}$, are shown in Fig.~\ref{fig:deadline:backlog},
and the radio footprints measured by average post-sparsification conflict degree $\mathbbm{E}_{\Omega}\left[d^s(v)\right] $ in Fig.~\ref{fig:deadline:degree}.

With zero-threshold policy (Dense-Zero), latency is low for low conflict densities $\bar{d}\leq 20$, but it increases quickly with $\bar{d}$ due to scheduling failures.
In comparison, non-zero threshold policies can retain more transmission opportunities, thus improving the latency when $\bar{d}>20$. 
In the target conflict densities ($60\leq \bar{d}\leq 100$), the global threshold policies Sparse-Stat and Sparse-Stat-(1) have the worst latency performances. 
The GCN-based threshold policies (Sparse-GCN($L$)) can further reduce the backlog over global threshold policies in the target densities, at the cost of slightly worse latency than the global threshold policies for $\bar{d}\leq 20$. 
Among GCN-based threshold policies, the latency of GCN(2) is slightly better than those of GCN(1) and GCN(3) within the target densities.
By leveraging network locality, Sparse-GCN(1) performs better than Sparse-Stat-(1), where the two have the same network-wide average thresholds.
Moreover, the hybrid threshold policy (Hybrid-GCN(1)) 
can achieve low latency across the full range of $\bar{d}$ by taking the best of Dense-Zero and Sparse-GCN(1), demonstrating the practicability of our approach.

Lastly, as shown in Fig.~\ref{fig:deadline:degree}, the average post-sparsification conflict degrees of GCN-based threshold policies are lower than those of the global threshold policies within the target densities, where GCN(1) achieves the lowest radio footprint for all $\bar{d}$. 
{Overall, GCN(1) can lead to the best latency and radio footprint for networks under LGS with fixed deadline.}

\begin{figure}[t]
	\centering
	\subfloat[]{
		\includegraphics[width=0.9\linewidth]{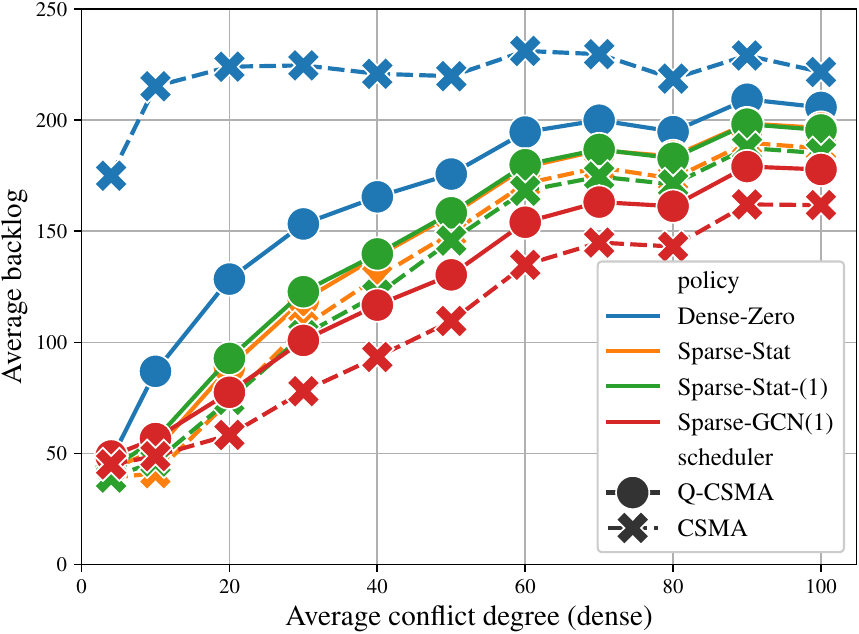}
		\label{fig:qcsma:backlog}
	}\\
	\subfloat[]{
		\includegraphics[width=0.9\linewidth]{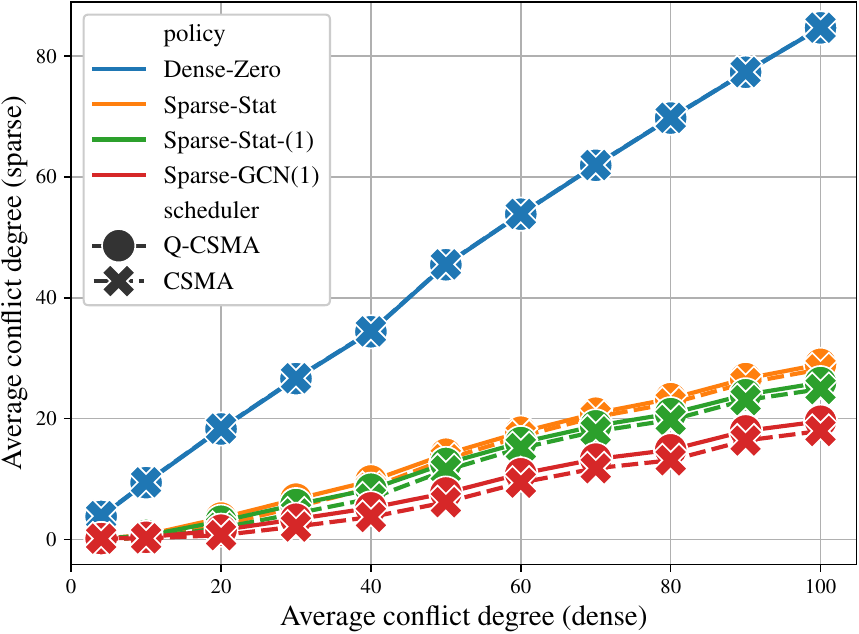}
		\label{fig:qcsma:degree}
	}
	\caption{The performance of threshold policies vs average conflict degree under Q-CSMA \cite{ni2012qcsma} and unweighted CSMA protocols specified in Section~\ref{sec:results:csma}:
	(a) average backlog, and
    (b) average post-sparsification conflict degree, averaged across time and network. 
	Each point is the average of 60-100 test instances. $\eta=0.75$.
	}   
	\label{fig:qcsma}    
\end{figure}

\subsubsection{LGS with Flexible Overhead}\label{sec:results:overhead}
In this experiment, the realized link rate is $r(v,t)\left[100-e(v,t)\right]/70$, where $e(v,t)$ is scheduling overhead of link $v$ at time step $t$ in ms,
the traffic setting and BA test set described in Section~\ref{sec:results:config} are utilized, with a global threshold $u^{(\eta)}=2310$ for $\eta=0.61$.
The average backlogs under tested threshold policies, as a function of the average conflict degree $\bar{d}$, are presented in Fig.~\ref{fig:overhead:backlog}, and the average post-sparsification conflict degrees in Fig.~\ref{fig:overhead:degree}.

With the same threshold policy, LGS with flexible overhead achieves lower latency compared to LGS with fixed deadline, as it compresses rather than discards a transmission opportunity when the scheduling overhead is high.
In the target densities ($60\leq \bar{d}\leq 100$), threshold policies based on deeper GCNs are ranked the best in both latency and radio footprint, followed by those with shallower GCNs, Sparse-Stat-(1), Sparse-Stat, and zero-threshold policy. 
The hybrid threshold policy (Hybrid-GCN(1)) outperforms both Sparse-GCN(1) and zero-threshold policy by taking the best of them, except that its radio footprint is higher than that of Sparse-GCN(1) for $\bar{d}<50$, which demonstrates its practical merit.
For networks under LGS with flexible overhead, deeper GCNs always translate to better latency and radio footprint.

\subsubsection{Q-CSMA and unweighted CSMA}\label{sec:results:csma}
Unweighted CSMA and Q-CSMA \cite{ni2012qcsma} both have a constant signaling overhead and no exponential backoff.
In the scheduling sub-slot of Q-CSMA, a link waits a random number of back-off mini-slots in $\left[0,W-1\right]$ before sending an INTENT message, and wins the contention if it sends the first INTENT message without collision.
The winner of the contention is scheduled with a probability of $e^{u(v)}/(e^{u(v)}+1)$.
A link scheduled in time-slot $t$ has priority to transmit in $t+1$, so that it can always deplete its backlog.
The unweighted CSMA shares the same procedure except that the winner of the contention is always scheduled.
We set the contention window $W=32$ and a mini-slot to be $1$ ms, thus a time-slot of 100 ms contains a scheduling sub-slot of 32 ms and a communication sub-slot of 68 ms.

In this experiment, the realized link rate equals the nominal one $r(v,t)$, the traffic setting and BA test set described in Section~\ref{sec:results:config} are utilized, with a global threshold of $u^{(\eta)}=5808$ corresponding to $\eta=0.75$.
The average backlogs under four threshold policies with Q-CSMA and unweighted CSMA, as a function of the average conflict degree, are presented in Fig.~\ref{fig:qcsma:backlog}, and the average post-sparsification conflict degrees in Fig.~\ref{fig:qcsma:degree}.
The GCN(1) model trained for LGS with fixed deadline in Section~\ref{sec:results:deadline} is reused for both Q-CSMA and unweighted CSMA.
For clarity of presentation, we do not show the results of Sparse-GCN(2) and GCN(3).

In the target densities ($60\leq\bar{d}\leq 100$), the latency of CSMA-based protocols is worse than those of LGS-based protocols due to increased scheduling failures caused by high collision rate in contention under a constant contention window $W$.
Under Q-CSMA, the largest improvements in latency and radio footprint over zero-threshold policy are brought by Sparse-GCN(1) for around $15\%$, followed by global threshold policies, Sparse-Stat-(1) and Sparse-Stat, which have similar latency but different radio footprints.
Although unweighted CSMA has the worst latency given zero-threshold policy, it significantly outperforms Q-CSMA with other non-zero threshold policies in both latency and radio footprint.
In particular, with Sparse-GCN(1), the latency of unweighted CSMA is $12\%$ lower than that of Q-CSMA. 
It suggests that unweighted CSMA with GCN-based threshold policy offers a better alternative to Q-CSMA.

\subsection{Discussions}\label{sec:results:discuss}

In Section~\ref{sec:results:graphs}, we demonstrated the advantages of GNNs and Alt-SGD over MLP and ZOO in training for signaling efficiency. 
In Section~\ref{sec:results:latency}, we showed that such efficiency can be translated to improved latency and radio footprint in wireless networks with different protocols.
In particular, the benefit of GNNs is two-fold: 1) global scaling on the average threshold in each network to compensate the mismatch between $\eta$ configured for a set of networks and the true $\eta^*$ for individual networks, evidenced by the gain of Sparse-Stat-(1) over Sparse-Stat; and 2) exploiting network locality, as shown by the gain of Sparse-GCN(1) over Sparse-Stat-(1).
In addition, a trained GCN can generalize to different topologies, and even from LGS to CSMA-based schedulers.
Notice that a $3:7$ split of scheduling-communication and a control message overhead as $1\%$ of frame length are chosen to keep simulations tractable on graphs of limited $\bar{d}$.
The demonstrated benefits extend to denser networks with other splits and smaller $\tau$.

\section{Conclusions}
\label{sec:conclusions}

We propose to reduce the scheduling overhead in dense wireless networks with delay-tolerant traffics, by withholding a link from scheduling contention if its utility is below a threshold.
The threshold of each link is individually adjusted by a GCN based on traffic statistics and network locality.  
Compared to no threshold or a statistical global threshold, GCN-based threshold policy can significantly reduce the overhead of MaxWeight scheduling and collision rate of CSMA in dense wireless networks, leading to improved latency, radio footprint or covertness, and energy efficiency. 
Moreover, GCNs can be implemented in a fully distributed, asynchronous manner and can generalize well across network topologies and even scheduling protocols. 
Our 1-layer GCN also has low communication and computing costs thanks to its  lightweight neural networks and message passing, and
thus can be deployed to low-end wireless devices in MTC and IoT systems.
Potential future directions include developing: 
1) efficient simulation techniques for large dense networks with higher sparsification ratio and/or random access protocols,
2) online learning approach to replace the offline search of statistical global threshold, and
3) solutions {for fairness and tail-delay metrics, as well as for scenarios with higher network mobility and dynamic power-control}.

\appendices

\begin{table}[t!]
	\renewcommand{\arraystretch}{1.05}
	\caption{Description of Major Notations
	} 
	\label{tab:symbols}
	\centering
    \setlength{\tabcolsep}{4pt}
	\footnotesize
	\begin{tabular}{|p{1.4cm}|p{6.8cm}|}
    \hline
		\textbf{Symbols}  & \textbf{Descriptions}   \\ \hline
    $ \alpha,\beta $ & $\alpha$: learning rate, $\beta$: decaying rate of $\alpha$ \\ \hline
    $\eta$ & global cut-off quantile $0\leq \eta \leq 1$ \\ \hline 
    $\tau$ & fraction of a time slot or frame duration for a link to broadcast a control message to its interfering neighbors \\ \hline 
    $\bbTheta_{0}^{l},\bbTheta_{1}^{l}$ & sets of trainable parameters of layer $l$ of the GCN in \eqref{E:gcn}  \\ \hline
    $ \lambda, \mu $ & $\lambda$ arrival rate, $\mu=\mathbb{E}(r)/\lambda$: traffic load \\ \hline
    $ \kappa_i, \rho_i $ & $\kappa_i$ ($\rho_i$) set of outgoing (incoming) links for device $i$ \\ \hline
    $\sigma_{l}(\cdot)$ & element-wise activation function for layer $l$ of GCN  \\ \hline
    $\bbPsi_{\ccalG}(\cdot;\mathbf{\bbomega})$ & GCN defined on graph $\ccalG$ with trainable parameters $\mathbf{\bbomega}$  \\ \hline
    $ \Omega$ & distribution of network state $(\ccalG,\bbu)$ \\ \hline
    $a(v,t)$ & No. of exogenous packets arriving link $v$ in time slot $t$ \\ \hline
    $c(\cdot)$ & function of the contention process \\ \hline
    $d(v), d^s(v)\newline \bar{d}$ & $d(v)$: degree of $v$, $d^s(v)$: post-sparsification degree of $v$,  \newline $\bar{d}$ average vertex degree of graph $\ccalG$  \\ \hline
    $D$ & minimal conflict degree to enable threshold in hybrid policy \\ \hline
    $f_c(\cdot)\newline f_s(\cdot)$ & function of collision rate\newline function of signaling overhead \\ \hline
    $f_{u}(\cdot), f'_{u}(\cdot)$ & the CDF and PDF of per-link utility \\ \hline
    $ f_{\mathbb{E}(u)}(\cdot) $ & function of the expected utility w.r.t. the cut-off probability for an individual link \\ \hline
    $g_{l}$ & dimension of output features of layer $l$ of GCN \\ \hline
    $\ccalG=(\ccalV,\ccalE)$ & conflict graph $\ccalG$, composed of a set of vertices $\ccalV$ and a set of edges $\ccalE$ \\ \hline
    $\ccalG^s\!\!=\!\!(\!\ccalV^s\!,\ccalE^s\!)\!$ & sparsified conflict graph $\ccalG^s$, composed of a set of vertices $\ccalV^s$ and a set of edges $\ccalE^s$ \\ \hline
    $h_v(\cdot)$ & local threshold function on vertex $v$ \\ \hline
    $H(\cdot)$ & Heaviside step function \\ \hline
    $K$ & The maximum number of control message broadcasting in LGS with fixed scheduling deadline \\ \hline
    $\ccalL$ & normalized Laplacian matrix of graph $\ccalG$  \\ \hline
    $ N $ & global norm clipping value in Algorithm~\ref{algo:foo} \\ \hline
    $\mathbb{N}(a,b)$ & normal distribution with mean $a$ \& standard deviation $b$  \\ \hline
    $\ccalN_{\ccalG}(v)$ &  set of immediate neighbors of vertex $v$ on graph $\ccalG$  \\ \hline
    $p(v)$ & cut off probability of link $v$ in scheduling \\ \hline
    $q(v,t), \newline q(v),\bbq_{v}\newline \bbq $ & $q(v,t)$: the queue length of link $v$ at time step $t$ \newline $q(v)=\bbq_{v}$: the queue length on link $v$ with $t$ omitted \newline $\bbq$: the vector of queue lengths on all links $\ccalV$ \\ \hline
    $r(v,t),\newline r(v),\bbr_{v}, \newline \bbr$ & $r(v,t)$: link rate of link $v$ at time step $t$ \newline $r(v)=\bbr_{v}$: link rate of link $v$ with $t$ omitted \newline $\bbr$: the vector of link rates for all links $\ccalV$  \\ \hline
    $u^{(\eta)}$ & global threshold for link utility  \\ \hline
    $u(v),\bbu_v\newline u(\boldsymbol{v})$ & $\bbu_{v}=u(v)$:  utility value on link $v$ \newline  total utility on independent set $\boldsymbol{v}$  \\ \hline
    $\bbu $ & $\bbu$: the vector of utility values on all links $\ccalV$  \\ \hline
    $ \mathbb{U}(a,b) $ & uniform distribution between $a$ and $b$ \\ \hline 
    $v$ \newline $\boldsymbol{v}$ &  $v\in\ccalV$: a vertex  (link) on the conflict graph $\ccalG$ (network) \newline $\boldsymbol{v}\subseteq\ccalV$: an  independent set on graph $\ccalG$  \\ \hline
    $\hat{\boldsymbol{v}}$ \newline $\hat{\boldsymbol{v}}^{s}\newline \tilde{\boldsymbol{v}}^{s}$ &   schedule from zero-threshold policy \newline  schedule from GCN-based threshold policy\newline  schedule from a reference threshold policy  \\ \hline
    $\bbv\newline \hat{\bbv}^s$ &  $\bbv\in\{0,1\}^{|\ccalV|}$: indicator vector of schedule $\boldsymbol{v}$ w.r.t. $\ccalV$\newline $ \hat{\bbv}^s\in\{0,1\}^{|\ccalV|} $: indicator vector of schedule $ \hat{\boldsymbol{v}^s} $ w.r.t. $\ccalV$  \\ \hline
    $\bbw^s$ & per-link utility vector under GCN-based threshold policy \\ \hline
    $\tilde{\bbw}^s$ & per-link utility vector under a reference threshold policy\;\; \\ \hline
    $\bbz,\bbz(\ccalG)$ & $\bbz\in\reals^{|\ccalV|}$: node embeddings of $\ccalG$ generated by GCN \\ \hline 
	\end{tabular}
\end{table}

\section{Computation of Partial Derivative in \eqref{eq:grad:e}}
\label{apx:grad:e}

By substituting $d^s(v)$ from \eqref{eq:spdeg} into \eqref{eq:obj:e}, we get
\begin{align*}
     & \mathbb{E}_{\Omega_{\bbu}^{\ccalG}} \left( |\ccalE^s | \big| \ccalG \right)\\
    =& \frac{1}{2} \sum_{v\in\ccalV}\left[d(v)-d(v)p(v)-\!\!\!\!\sum_{i\in\ccalN_{\ccalG}(v)}\!\!p(i)+\!\!\!\!\sum_{i\in\ccalN_{\ccalG}(v)}\!\!p(v)p(i)\right]\\
    =& \frac{1}{2}\left[\sum_{v\in\ccalV}d(v)-2\sum_{v\in\ccalV}d(v)p(v)+\sum_{v\in\ccalV}\sum_{i\in\ccalN_{\ccalG}(v)}p(v)p(i)\right]\;.
\end{align*}
This equation holds because  $$ \sum_{v\in\ccalV}\sum_{i\in\ccalN_{\ccalG}(v)}p(i)= \sum_{v\in\ccalV}d(v)p(v),$$ where each node is counted once by each of its neighbors in the left hand side.
For each neighbor $i\in\ccalN_{\ccalG}(v)$ of a particular $v$, $p(v)p(i)$ is counted twice  in $ \sum_{v\in\ccalV}\sum_{i\in\ccalN_{\ccalG}(v)}p(v)p(i) $. 
Therefore, the partial derivative of \eqref{eq:obj:e} w.r.t. $p(v)$ is $$ \frac{\partial \mathbb{E}_{\Omega_{\bbu}^{\ccalG}} \left( |\ccalE^s | \big| \ccalG \right)}{\partial p(v)}=-d(v)+\sum_{i\in\ccalN_{\ccalG}(v)}p(i)=-d^s(v). $$ 
where we used \eqref{eq:spdeg}. 
Based on chain rule, we have 
\begin{equation*}
    \frac{\partial \mathbb{E}_{\Omega_{\bbu}^{\ccalG}} \left( |\ccalE^s | \big| \ccalG \right)}{\partial z(v)} =  -d^s(v)f_u'\left(z(v)u^{(\eta)}\right)u^{(\eta)}\;.
\end{equation*}

\section{}
\label{apx:concave}
We solve~\eqref{eq:eta} with Algo.~\ref{algo:search}, where $f_{\eta}(\cdot)$ is the objective in~\eqref{eq:eta}, $a=0, b=1$.
It has a complexity of $\ccalO(\log(\frac{b-a}{\epsilon}))$, as the search interval is cut by $1/2$ or $1/4$ in each iteration.

\begin{algorithm}[h]
\caption{Peak Search for Quasi-concave Function}
\label{algo:search}
\hspace*{\algorithmicindent} 
\textbf{Input}: $f_{\eta}(\cdot), [a,b], 0<\epsilon\ll 1$ \\
\hspace*{\algorithmicindent} 
\textbf{Output}: $x_m$ \COMMENTS{The near optimal, $|\eta^* -x_m|\leq \epsilon $}\\ \vspace{-0.2in}
\begin{algorithmic}[1] 
\STATE $x_{l}=a, x_{r}=b, x_m=(x_l+x_r)/2$
\WHILE{ $ x_{r}-x_{l} > \epsilon $}
\STATE $x_1 = \argmax\limits_{x\in\{x_{l}, x_{r}, x_{m}\}} \!\!f_{\eta}(x)$, $x_2=\argmax\limits_{x\in\{x_{l}, x_{r}, x_{m}\}\setminus \{x_1\} }\!\!\!f_{\eta}(x)$
\STATE $x_3=(x_1 + x_2)/2$
\IF{ $f_{\eta}(x_3) \geq f_{\eta}(x_1) $ }
\STATE $ x_m \leftarrow x_3 $, $x_l\leftarrow \min(x_1, x_2)$, $x_r\leftarrow \max(x_1, x_2)$
\ELSE
\STATE $ x_m \leftarrow x_1 $, $x_l\leftarrow \min(x_3, x_l)$, $x_r\leftarrow \max(x_3, x_r)$
\ENDIF
\ENDWHILE
\end{algorithmic}
\end{algorithm}


\bibliographystyle{ieeetr}
\bibliography{strings,refs}

\begin{thebibliography}{10}

\bibitem{zhao2022icassp_b}
Z.~Zhao, A.~Swami, and S.~Segarra, ``Distributed link sparsification for
  scalable scheduling using graph neural networks,'' in {\em IEEE Intl. Conf.
  Acoustics, Speech and Signal Process. (ICASSP)}, pp.~5308--5312, 2022.

\bibitem{cisco2020}
``Cisco annual internet report (2018–2023),'' white paper, Cisco Systems,
  Inc., Mar. 2020.
\newblock [Online]. Available:
  \url{https://www.cisco.com/c/en/us/solutions/collateral/executive-perspectives/annual-internet-report/white-paper-c11-741490.html}.

\bibitem{kott2016internet}
A.~Kott, A.~Swami, and B.~J. West, ``The internet of battle things,'' {\em
  Computer}, vol.~49, no.~12, pp.~70--75, 2016.

\bibitem{akyildiz20206g}
I.~F. Akyildiz, A.~Kak, and S.~Nie, ``{6G} and beyond: The future of wireless
  communications systems,'' {\em IEEE Access}, vol.~8, pp.~133995--134030,
  2020.

\bibitem{chen2021massive}
X.~Chen, D.~W.~K. Ng, W.~Yu, E.~G. Larsson, N.~Al-Dhahir, and R.~Schober,
  ``Massive access for {5G} and beyond,'' {\em IEEE J. Sel. Areas Commun.},
  vol.~39, no.~3, pp.~615--637, 2021.

\bibitem{Sharma2020toward}
S.~K. Sharma and X.~Wang, ``Toward massive machine type communications in
  ultra-dense cellular {IoT} networks: Current issues and machine
  learning-assisted solutions,'' {\em IEEE Commun. Surveys \& Tutorials},
  vol.~22, no.~1, pp.~426--471, 2020.

\bibitem{Lin06}
X.~Lin, N.~B. {Shroff}, and R.~{Srikant}, ``A tutorial on cross-layer
  optimization in wireless networks,'' {\em IEEE J. Sel. Areas Commun.},
  vol.~24, no.~8, pp.~1452--1463, 2006.

\bibitem{sarkar2013ad}
S.~K. Sarkar, T.~G. Basavaraju, and C.~Puttamadappa, {\em Ad Hoc Mobile
  Wireless Networks: Principles, Protocols and Applications}.
\newblock Boca Raton, FL, USA: CRC Press, Taylor \& Francis Group, 2nd~ed.,
  2013.

\bibitem{Patriciello2016}
N.~Patriciello, C.~A. Grazia, J.~Núñez-Martínez, J.~Baranda,
  J.~Mangues-Bafalluy, and M.~Casoni, ``Performance evaluation of backpressure
  routing in integrated satellite-terrestrial backhaul for {PPDR} networks,''
  in {\em Proc. IEEE Intl. Conf. Wireless Mobile Comput., Netw., Commun.
  (WiMob)}, pp.~1--8, 2016.

\bibitem{Cudak2021}
M.~Cudak, A.~Ghosh, A.~Ghosh, and J.~Andrews, ``Integrated access and backhaul:
  A key enabler for {5G} millimeter-wave deployments,'' {\em IEEE Commun.
  Mag.}, vol.~59, no.~4, pp.~88--94, 2021.

\bibitem{zhao2023iclr}
Z.~Zhao, A.~Swami, and S.~Segarra, ``Graph-based deterministic policy gradient
  for repetitive combinatorial optimization problems,'' in {\em Intl. Conf.
  Learn. Repres. (ICLR)}, pp.~1--21, 2023.

\bibitem{zhao2022twc}
Z.~Zhao, G.~Verma, C.~Rao, A.~Swami, and S.~Segarra, ``Link scheduling using
  graph neural networks,'' {\em IEEE Trans. Wireless Commun.}, vol.~22, no.~6,
  pp.~3997--4012, 2023.

\bibitem{Joo09}
C.~{Joo}, X.~{Lin}, and N.~B. {Shroff}, ``Understanding the capacity region of
  the greedy maximal scheduling algorithm in multihop wireless networks,'' {\em
  IEEE/ACM Trans. Netw.}, vol.~17, no.~4, pp.~1132--1145, 2009.

\bibitem{marques2011optimal}
A.~G. Marques, N.~Gatsis, and G.~B. Giannakis, ``Optimal cross-layer design of
  wireless fading multi-hop networks,'' in {\em Cross Layer Designs in WLAN
  Systems} (N.~Zorba, C.~Skianis, and C.~Verikoukis, eds.), pp.~1--44,
  Leicester, U.K.: Troubador Publishing Ltd., 2011.

\bibitem{basagni2001finding}
S.~Basagni, ``Finding a maximal weighted independent set in wireless
  networks,'' {\em Telecommun. Systems}, vol.~18, no.~1-3, pp.~155--168, 2001.

\bibitem{cheng2009complexity}
W.~Cheng, X.~Cheng, T.~Znati, X.~Lu, and Z.~Lu, ``The complexity of channel
  scheduling in multi-radio multi-channel wireless networks,'' in {\em IEEE
  Intl. Conf. Computer Comms. (INFOCOM)}, pp.~1512--1520, 2009.

\bibitem{joo2012local}
C.~{Joo} and N.~B. {Shroff}, ``Local greedy approximation for scheduling in
  multihop wireless networks,'' {\em IEEE Trans. Mobile Computing}, vol.~11,
  no.~3, pp.~414--426, 2012.

\bibitem{joo2015distributed}
C.~Joo, X.~Lin, J.~Ryu, and N.~B. Shroff, ``Distributed greedy approximation to
  maximum weighted independent set for scheduling with fading channels,'' {\em
  IEEE/ACM Trans. Netw.}, vol.~24, no.~3, pp.~1476--1488, 2015.

\bibitem{du2016new}
P.~Du and Y.~Zhang, ``A new distributed approximation algorithm for the maximum
  weight independent set problem,'' {\em Mathematical Problems in Engineering},
  vol.~2016, no.~1, p.~9790629, 2016.

\bibitem{paschalidis2015message}
I.~C. Paschalidis, F.~Huang, and W.~Lai, ``A message-passing algorithm for
  wireless network scheduling,'' {\em IEEE/ACM Trans. Netw.}, vol.~23,
  pp.~1528--1541, Oct. 2015.

\bibitem{joo2010complexity}
C.~Joo, G.~Sharma, N.~B. Shroff, and R.~R. Mazumdar, ``On the complexity of
  scheduling in wireless networks,'' {\em EURASIP J. Wireless Commun. Netw.},
  vol.~2010, no.~1, p.~418934, 2010.

\bibitem{zhao2021icassp}
Z.~Zhao, G.~Verma, C.~Rao, A.~Swami, and S.~Segarra, ``Distributed scheduling
  using graph neural networks,'' in {\em IEEE Intl. Conf. Acoustics, Speech and
  Signal Process. (ICASSP)}, pp.~4720--4724, 2021.

\bibitem{ni2012qcsma}
J.~Ni, B.~Tan, and R.~Srikant, ``{Q-CSMA}: Queue-length-based {CSMA/CA}
  algorithms for achieving maximum throughput and low delay in wireless
  networks,'' {\em IEEE/ACM Trans. Netw.}, vol.~20, no.~3, pp.~825--836, 2012.

\bibitem{jiang2010distCSMA}
L.~Jiang and J.~Walrand, ``A distributed {CSMA} algorithm for throughput and
  utility maximization in wireless networks,'' {\em IEEE/ACM Trans. Netw.},
  vol.~18, no.~3, pp.~960--972, 2010.

\bibitem{Lin2009constant}
X.~Lin and S.~B. Rasool, ``Constant-time distributed scheduling policies for ad
  hoc wireless networks,'' {\em IEEE Trans. Auto. Control}, vol.~54, no.~2,
  pp.~231--242, 2009.

\bibitem{luby1985simple}
M.~Luby, ``A simple parallel algorithm for the maximal independent set
  problem,'' in {\em Proc. 17th annual ACM Symp. Theory of computing},
  pp.~1--10, 1985.

\bibitem{Testi2021blind}
E.~Testi and A.~Giorgetti, ``Blind wireless network topology inference,'' {\em
  IEEE Trans. Commun.}, vol.~69, no.~2, pp.~1109--1120, 2021.

\bibitem{ye2004coordinated}
W.~Ye, J.~Heidemann, and D.~Estrin, ``Medium access control with coordinated
  adaptive sleeping for wireless sensor networks,'' {\em IEEE/ACM Trans.
  Netw.}, vol.~12, no.~3, pp.~493--506, 2004.

\bibitem{Santi2005topology}
P.~Santi, ``Topology control in wireless ad hoc and sensor networks,'' {\em ACM
  Comput. Surv.}, vol.~37, pp.~164--194, June 2005.

\bibitem{Ramanathan2004ch5}
R.~Ramanathan, ``Antenna beamforming and power control for ad hoc networks,''
  in {\em Mobile Ad Hoc Networking} (S.~Basagni, M.~Conti, S.~Giordano, and
  I.~Stojmenovic, eds.), pp.~139--173, Hoboken, NJ, USA: John Wiley \& Sons,
  Ltd, 2004.

\bibitem{ray2016hybrid}
N.~K. Ray and A.~K. Turuk, ``A hybrid energy efficient protocol for mobile ad
  hoc networks,'' {\em J. Comput. Netw. Commun.}, vol.~2016, no.~1, p.~2861904,
  2016.

\bibitem{guha2011greenwave}
S.~Guha, P.~B. Basu, C.-K.~C. Chau, and R.~Gibbens, ``Green wave sleep
  scheduling: Optimizing latency and throughput in duty cycling wireless
  networks,'' {\em IEEE J. Sel. Areas Commun.}, vol.~29, no.~8, pp.~1595--1604,
  2011.

\bibitem{long2020collaborative}
J.~Long and O.~B\"{u}y\"{u}k\"{o}zt\"{u}rk, ``Collaborative duty cycling
  strategies in energy harvesting sensor networks,'' {\em Computer-Aided Civil
  and Infrastructure Engineering}, vol.~35, pp.~534--548, May 2020.

\bibitem{lin2010lowcomplexity}
L.~Lin, X.~Lin, and N.~B. Shroff, ``Low-complexity and distributed energy
  minimization in multihop wireless networks,'' {\em IEEE/ACM Trans. Netw.},
  vol.~18, no.~2, pp.~501--514, 2010.

\bibitem{xiang2014energy}
X.~Xiang, C.~Lin, and X.~Chen, ``Energy-efficient link selection and
  transmission scheduling in mobile cloud computing,'' {\em IEEE Commun.
  Lett.}, vol.~3, no.~2, pp.~153--156, 2014.

\bibitem{wu2020energy}
W.~Wu, W.~Xu, Z.~Chen, and M.~Yang, ``Energy-efficient link scheduling in
  time-variant dual-hop {60GHz} wireless networks,'' {\em Concurrency and
  Computation: Practice and Experience}, vol.~32, no.~23, p.~e5903, 2020.

\bibitem{Ahlswede2000}
R.~Ahlswede, N.~Cai, S.-Y. Li, and R.~Yeung, ``Network information flow,'' {\em
  IEEE Trans. Info. Theory}, vol.~46, no.~4, pp.~1204--1216, 2000.

\bibitem{hai2018delay}
L.~Hai, Q.~Gao, J.~Wang, H.~Zhuang, and P.~Wang, ``Delay-optimal back-pressure
  routing algorithm for multihop wireless networks,'' {\em IEEE Trans.
  Vehicular Tech.}, vol.~67, no.~3, pp.~2617--2630, 2018.

\bibitem{wang2024generalization}
Z.~Wang, J.~Cervi{\~n}o, and A.~Ribeiro, ``Generalization of graph neural
  networks is robust to model mismatch,'' in {\em Proc. AAAI Conf. Artificial
  Intell. (AAAI)}, (Philadelphia, PA, USA), pp.~2387--2395, AAAI Press, 2025.

\bibitem{wang2022learning}
Z.~Wang, M.~Eisen, and A.~Ribeiro, ``Learning decentralized wireless resource
  allocations with graph neural networks,'' {\em IEEE Trans. Signal Process.},
  vol.~70, pp.~1850--1863, 2022.

\bibitem{liu2020primer}
S.~Liu, P.-Y. Chen, B.~Kailkhura, G.~Zhang, A.~O. Hero~III, and P.~K. Varshney,
  ``A primer on zeroth-order optimization in signal processing and machine
  learning: Principals, recent advances, and applications,'' {\em IEEE Signal
  Process. Mag.}, vol.~37, no.~5, pp.~43--54, 2020.

\bibitem{huang2012low}
P.-K. Huang, X.~Lin, and C.-C. Wang, ``A low-complexity congestion control and
  scheduling algorithm for multihop wireless networks with order-optimal
  per-flow delay,'' {\em IEEE/ACM Trans. Netw.}, vol.~21, no.~2, pp.~495--508,
  2012.

\bibitem{gupta2009low}
A.~Gupta, X.~Lin, and R.~Srikant, ``Low-complexity distributed scheduling
  algorithms for wireless networks,'' {\em IEEE/ACM Trans. Netw.}, vol.~17,
  no.~6, pp.~1846--1859, 2009.

\bibitem{Ryu2012timescale}
J.~Ryu, L.~Ying, and S.~Shakkottai, ``Timescale decoupled routing and rate
  control in intermittently connected networks,'' {\em IEEE/ACM Trans. Netw.},
  vol.~20, no.~4, pp.~1138--1151, 2012.

\bibitem{yadin1963queueing}
M.~Yadin and P.~Naor, ``Queueing systems with a removable service station,''
  {\em J. Operational Research Soc.}, vol.~14, pp.~393--405, 1963.

\bibitem{kella1989threshold}
O.~Kella, ``The threshold policy in the {M/G/1} queue with server vacations,''
  {\em Naval Research Logistics}, vol.~36, no.~1, pp.~111--123, 1989.

\bibitem{lee2013n}
D.~H. Lee and W.~S. Yang, ``The n-policy of a discrete time {Geo/G/1} queue
  with disasters and its application to wireless sensor networks,'' {\em
  Applied Mathematical Modelling}, vol.~37, no.~23, pp.~9722--9731, 2013.

\bibitem{Qi2020traffic}
N.~Qi, N.~I. Miridakis, M.~Xiao, T.~A. Tsiftsis, R.~Yao, and S.~Jin,
  ``Traffic-aware two-stage queueing communication networks: Queue analysis and
  energy saving,'' {\em IEEE Trans. Commun.}, vol.~68, no.~8, pp.~4919--4932,
  2020.

\bibitem{wu2021tnnls}
Z.~Wu, S.~Pan, F.~Chen, G.~Long, C.~Zhang, and P.~S. Yu, ``A comprehensive
  survey on graph neural networks,'' {\em IEEE Trans. Neural Netw. Learn.
  Syst.}, vol.~32, no.~1, pp.~4--24, 2021.

\bibitem{busoniu2008comprehensive}
L.~Busoniu, R.~Babuska, and B.~De~Schutter, ``A comprehensive survey of
  multiagent reinforcement learning,'' {\em IEEE Trans. Systems, Man,
  Cybernetics, Part C (Appl. Rev.)}, vol.~38, no.~2, pp.~156--172, 2008.

\bibitem{giselsson2018large}
P.~Giselsson and A.~Rantzer, {\em Large-scale and Distributed Optimization},
  vol.~2227.
\newblock Cham, Switzerland: Springer, 2018.

\bibitem{ross2011reduction}
S.~Ross, G.~Gordon, and D.~Bagnell, ``A reduction of imitation learning and
  structured prediction to no-regret online learning,'' in {\em Proc. 14th
  Intl. Conf. Artificial Intelligence and Statistics}, pp.~627--635, JMLR
  Wrksp. \& Conf. Proc., 2011.

\bibitem{watkins1992q}
C.~J. Watkins and P.~Dayan, ``Q-learning,'' {\em Machine learning}, vol.~8,
  no.~3-4, pp.~279--292, 1992.

\bibitem{silver2016mastering}
D.~Silver, A.~Huang, C.~J. Maddison, A.~Guez, L.~Sifre, G.~Van Den~Driessche,
  J.~Schrittwieser, I.~Antonoglou, V.~Panneershelvam, M.~Lanctot, {\em et~al.},
  ``Mastering the game of go with deep neural networks and tree search,'' {\em
  Nature}, vol.~529, no.~7587, pp.~484--489, 2016.

\bibitem{silver2014deterministic}
D.~Silver, G.~Lever, N.~Heess, T.~Degris, D.~Wierstra, and M.~Riedmiller,
  ``Deterministic policy gradient algorithms,'' in {\em Intl. Conf. Machine
  Learning (ICML)}, pp.~387--395, PMLR, 2014.

\bibitem{barenboim2013distributed}
L.~Barenboim and M.~Elkin, ``Distributed graph coloring: Fundamentals and
  recent developments,'' {\em Synthesis Lectures on Distributed Computing
  Theory}, vol.~4, no.~1, pp.~1--171, 2013.

\bibitem{Mousavi17lte}
H.~Mousavi, I.~S. Amiri, M.~Mostafavi, and C.~Choon, ``{LTE} physical layer:
  Performance analysis and evaluation,'' {\em Applied Computing and
  Informatics}, vol.~15, no.~1, pp.~34 -- 44, 2019.

\bibitem{erdds1959random}
P.~Erdős and A.~Rényi, ``On random graphs {I},'' {\em Publ. Math. Debrecen
  6}, pp.~290--297, 1959.

\bibitem{barabasi2016network}
A.-L. Barab\'asi and M.~P\'osfai, {\em Network Science}.
\newblock Cambridge, U.K.: Cambridge Univ. Press, 2016.

\end{thebibliography}

\begin{IEEEbiography}
[{\includegraphics[width=1in,height=1.25in,clip,keepaspectratio]{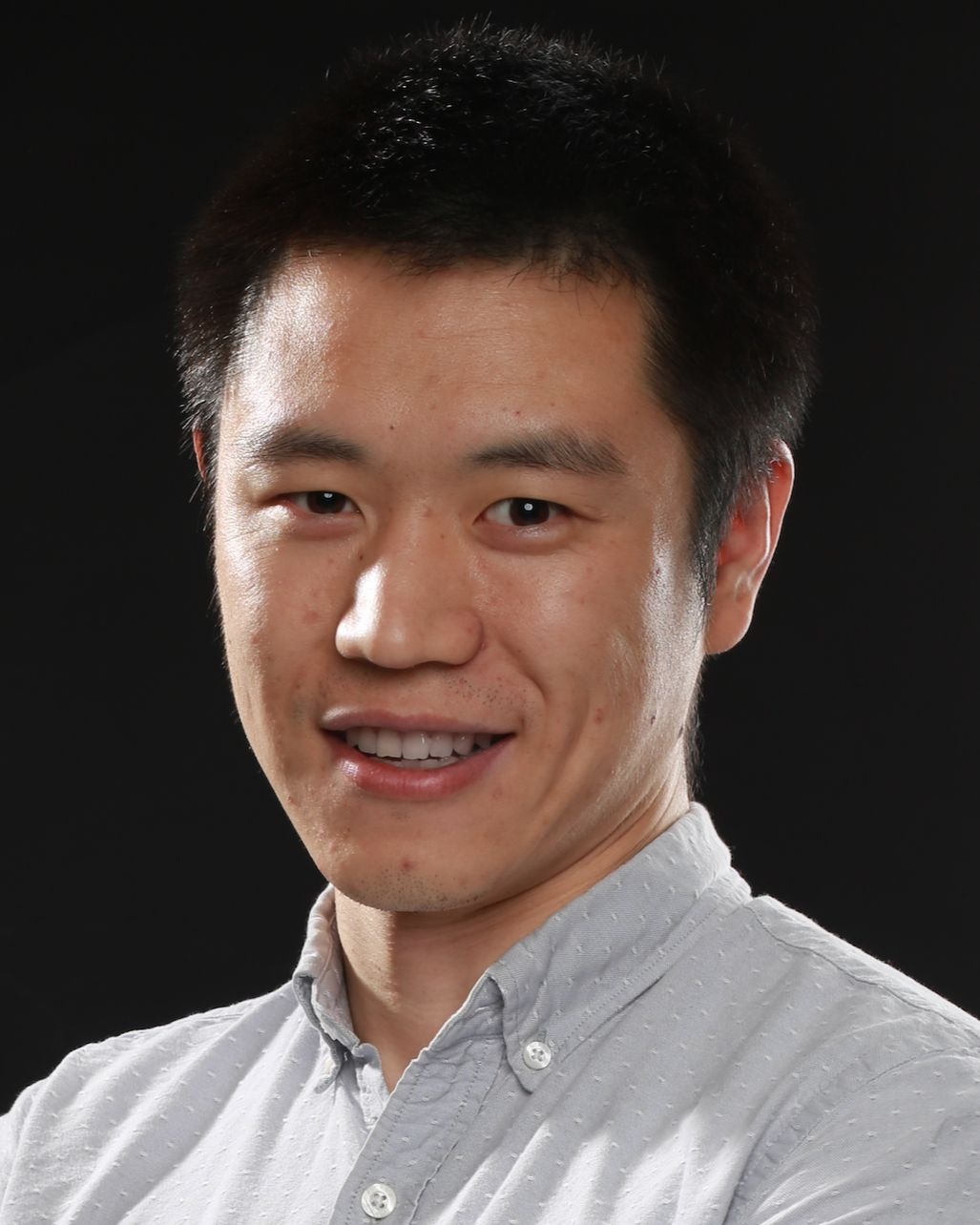}}]{Zhongyuan Zhao} (Member, IEEE) received his B.Sc. and M.S. degrees in Electronic Engineering from the University of Electronic Science and Technology of China, Chengdu, China, in 2006 and 2009, respectively. He received his Ph.D. degree in Computer Engineering from the University of Nebraska-Lincoln, Lincoln, NE, in 2019, under the guidance of Prof. Mehmet C. Vuran. From 2009 to 2013, he worked for ArrayComm and Ericsson, respectively, as an engineer developing 4G base-stations. He joined Rice University in 2019 as a postdoctoral research associate at the Department of Electrical and Computer Engineering, advised by Prof. Santiago Segarra. Currently, he is a Research Assistant Professor at Rice University. Dr. Zhao’s current research interests include machine learning, signal processing, and dynamic resource allocation in wireless communications and networked systems.
\end{IEEEbiography}
\vspace{-0.5in}

\begin{IEEEbiography}
[{\includegraphics[width=1in,height=1.25in,clip,keepaspectratio]{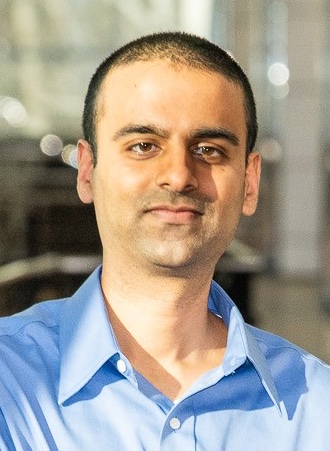}}]{Gunjan Verma}
received the B.S. degree in mathematics, computer science, and economics from Rutgers University, the M.A. degree in computational biology from Duke University, and the M.S. degree in mathematics from Johns Hopkins University. He is currently a Senior Computer Scientist and a Research and Development Portfolio Director with the U.S. Army Research Laboratory (ARL), Adelphi, MD, USA. 
In 2025, Mr. Verma received the Civilian Service Commendation Medal from the U.S. Department of the Army in recognition of his exceptional leadership, research, and development efforts in several high-impact research and development initiatives.
His research interests include Bayesian statistics and machine learning and their application to problems in networking, communications, and robotics.
\end{IEEEbiography}
 \vspace{-0.5in}


\begin{IEEEbiography}
[{\includegraphics[width=1in,height=1.25in,clip,keepaspectratio]{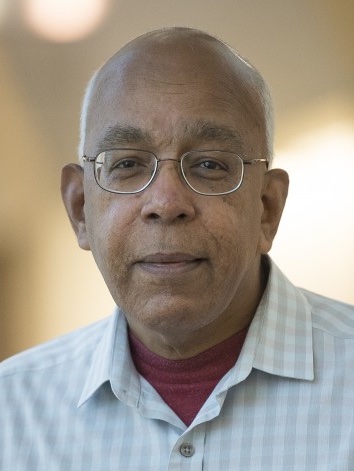}}]{Ananthram Swami}
(Life Fellow, IEEE) received the B.Tech. degree from IIT-Bombay, the M.S. degree from Rice University, and the Ph.D. degree from the University of Southern California (USC), all in electrical engineering. He is currently with the U.S. Army's DEVCOM Army Research Laboratory (ARL) as the Army’s Senior Research Scientist (ST) for Network Science, and is an ARL fellow. Prior to joining ARL, he held positions with Unocal Corporation, USC, and CS-3. He has held visiting faculty positions at INP, Toulouse, and Imperial College, London. 
He is a co-editor of the 2023 IEEE/Wiley book {\it IoT for Defense and National Security}, and coauthor of the
2021 Cambridge University Press book {\it Network Tomography}.
His recent awards include 2024 IEEE LANMAN and 2024 IEEE ICC best paper awards, 2018 IEEE ComSoc MILCOM Technical Achievement Award and the 2017 Presidential Rank Award (Meritorious).
\end{IEEEbiography}
 \vspace{-0.5in}

\begin{IEEEbiography}
[{\includegraphics[width=1in,height=1.25in,clip,keepaspectratio]{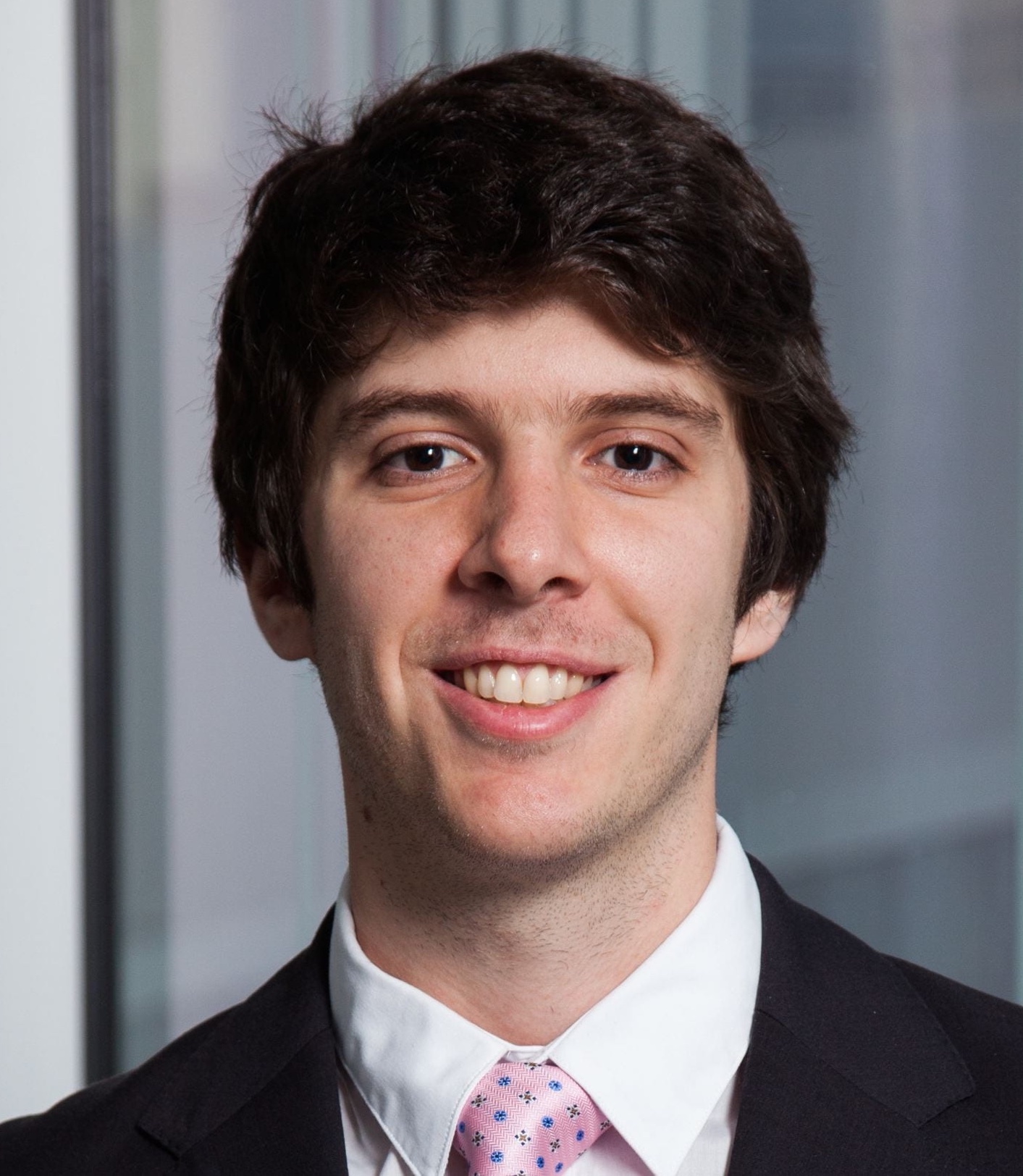}}]{Santiago Segarra} (Senior Member, IEEE) received the B.Sc. degree (Hons.) (Valedictorian) in industrial engineering from the Instituto Tecnológico de Buenos Aires (ITBA), Argentina, in 2011, the M.Sc. in electrical engineering from the University of Pennsylvania (Penn), Philadelphia, in 2014 and the Ph.D. degree in electrical and systems engineering from Penn in 2016. From September 2016 to June 2018 he was a postdoctoral research associate with the Institute for Data, Systems, and Society at the Massachusetts Institute of Technology. 
In 2018, Dr. Segarra joined the Department of Electrical and Computer Engineering, Rice University as an Assistant Professor, and is currently an Associate Professor. He also holds courtesy appointments with the Departments of Computer Science and Statistics. His research interests include network theory, data analysis, machine learning, and graph signal processing. He received the 2011 Outstanding Graduate Award granted by the National Academy of Engineering of Argentina, the 2017 Penn’s Joseph and Rosaline Wolf Award for Best Doctoral Dissertation in Electrical and Systems Engineering, the 2020 IEEE Signal Processing Society Young Author Best Paper Award, the 2021 Rice’s School of Engineering Research + Teaching Excellence Award, three early career awards (NSF CAREER, ARO ECP, and ARI Early Career), and five best conference paper awards. Since 2022, he has dedicated part of his time consulting for Microsoft Research.
\end{IEEEbiography}

\end{document}